\documentclass[10pt,preprintnumbers,amsmath,amssymb,nofootinbib,superscriptaddress, floatfix]{revtex4-2}

\usepackage[english]{babel}

\usepackage[utf8]{inputenc}
\usepackage[T1]{fontenc}
\usepackage{ucs}
\usepackage{microtype}
\usepackage{newtxtext,newtxmath}
\usepackage{color}
\usepackage{lettrine}

\usepackage{url}
\usepackage{hyperref}

\usepackage{xspace}
\usepackage{lastpage}
\usepackage{stmaryrd}

\usepackage{graphicx}
\usepackage{empheq}
\usepackage{ifthen}
\usepackage{tensor}
\usepackage{paralist}
\usepackage{float}

\usepackage{slashed}
\usepackage{amsmath}
\usepackage{amsfonts}
\usepackage{amssymb}
\usepackage{mathrsfs}
\usepackage{extarrows}
\usepackage{verbatim}
\usepackage{upgreek}
\usepackage{wasysym}
\usepackage{multirow}

\usepackage{natbib}

\def \be {\begin{equation}}
\def \ee {\end{equation}}
\def \dd {\mathrm{d}} 
\def \t {\tilde}

\def \l {\left}
\def \r {\right}

\def \bs {\boldsymbol}

\def \T {\Theta}
\newcommand{\e}[1]{_{\rm #1}}

\newcommand{\beq}{\begin{equation}}
\newcommand{\eeq}{\end{equation}}
\newcommand{\bea}{\begin{eqnarray}}
\newcommand{\eea}{\end{eqnarray}}


\newcommand\ees{\end{eqnarray}}
\newcommand\bees{\begin{eqnarray}}

\begin{document}

\title{Peculiar velocity effects on the Hubble constant from time-delay cosmography}

\author{Charles Dalang}
\email{c.dalang@qmul.ac.uk}
\affiliation{Queen Mary University of London, Mile End Road, London E1 4NS, United Kingdom}
\author{Martin Millon}
\email{millon@stanford.edu}
\affiliation{Kavli Institute for Particle Astrophysics and Cosmology, Stanford University, Stanford, California 94305, USA}
\author{Tessa Baker}
\email{t.baker@qmul.ac.uk}
\affiliation{Queen Mary University of London, Mile End Road, London E1 4NS, United Kingdom}

\begin{abstract}
Two major challenges of contemporary cosmology are the Hubble tension and the cosmic dipole tension. At the crossroads of these, we investigate the impact of peculiar velocities on estimations of the Hubble constant from time-delay cosmography. We quantify the bias on the inference of the Hubble constant due to peculiar velocities of the lens, the source and of the observer. The former two, which may cancel from one system to another, affect the determination of the angular diameter distances in the time-delay formula, and reconstructed quantities like the angle to the source, via a lens model. On the other hand, the peculiar velocity of the observer, which is a debated quantity in the context of the cosmic dipole tension, systematically affects observed angles through aberration, redshifts, angular diameter distance and reconstructed quantities. We compute in detail the effect of these peculiar velocities on the inference of the Hubble constant to linear order in the peculiar velocities for the seven lenses of the H0LiCOW/TDCOSMO collaboration. The bias generated by the observer's peculiar velocity alone can reach $1.15\%$ for the lenses which are well aligned with it. This results in a $0.25 \%$ bias for the seven combined lenses. Assuming a typical peculiar velocity of $300$ km s$^{-1}$ for the lens and the source galaxies, these add an additional random uncertainty, which can reach $1\%$ for an individual lens, but reduces to $0.24\%$ for the full TDCOSMO sample. The picture may change if peculiar velocities turn out to be larger than expected. Any time-delay cosmography program which aims for percent precision on the Hubble constant may need to take this source of systematic bias into account. This is especially so for future ground-based surveys which cover a fraction of the celestial sphere that is well aligned with the observer's peculiar velocity.
\end{abstract}

\maketitle

\section{Introduction}

\lettrine{P}{}ersistent tensions in cosmological datasets may be indicators of new physics or of unknown systematics. While the former is very exciting, excluding confidently the latter is notoriously difficult. On the theoretical side, this is mostly because in extracting cosmological parameters, approximations are needed, which require a set of assumptions that may be broken. Two of these tensions include disagreement on the kinematic cosmic dipole, which can be translated into a tension on the peculiar velocity of the observer \cite{Aluri:2022hzs} and on the Hubble constant \cite{Di_Valentino_2021}. Both of these tensions are between the cosmic microwave background (CMB) and other datasets. 

The CMB dipole allows one to extract the velocity of the observer, which effectively Doppler shifts the black body radiation of angular average temperature $\langle T\rangle $ from the CMB to higher and lower temperatures $(\delta T/\langle T\rangle)\e{dip} \sim \mathcal{O}(10^{-3})$ in opposite hemispheres aligned with the observer's velocity. This works well provided the intrinsic CMB dipole, which is expected to be of the order $\mathcal{O}(10^{-5})$, is small in comparison. This is expected from a nearly scale-invariant power spectrum of primordial fluctuations of the inflaton generated at the end of a period of quasi-de Sitter expansion during inflation. Under the assumption that the intrinsic CMB dipole vanishes, known as the \textit{entirely kinematic} interpretation of the CMB dipole, one obtains $||\bs{v}\e{dip}||= 369.82\pm 0.11$ km\,s$^{-1}$ toward $\bs{\hat{v}}\e{dip} = (264.021^\circ \pm 0.011^\circ, 48.253^\circ \pm 0.005^\circ )$ in galactic coordinates \cite{Fixsen:1994, Fixsen:1996,Aghanim:2018eyx,Planck:2013kqc}. This defines a reference frame known as the CMB frame. If the interpretation is correct, the same velocity should induce correlations between the $l$ and $l\pm 1$ multipoles of the CMB, which was checked in \cite{Planck:2013kqc, Ferreira:2020aqa, Saha:2021bay} and gives consistent results, %\tb{,}
albeit the relatively large error bars still leave room for an intrinsic dipole which can make up to $40\%$ of the CMB dipole \cite{Schwarz:2016}. It should be noted that spectral distortions of the CMB monopole, dipole, and quadrupole should let one separate the intrinsic dipole from its kinematic counterpart with sufficiently advanced detectors \cite{Yasini:2016dnd}. 

Alternatively, the peculiar velocity of the observer can be extracted from source number counts of relatively high redshift sources ($z\geq 0.1$), such as quasars, to avoid contamination from local structures \cite{Tiwari_2016,Dalang:2021ruy}. This was pioneered by G.\,Ellis and J.\,Baldwin for flux-limited surveys of sources with a flux density following a power law frequency spectrum \cite{Ellis1984}. Aberration of angles and Doppler shift then affect these number counts per unit solid angle in such a way that permits the extraction of the observer's peculiar velocity with respect to these sources. This has led to a number count dipole, which is well aligned with the CMB dipole but about $2-5$ times as large as expected from $\bs{v}\e{dip}$ and which has reached a $\sim 5\sigma$ tension \cite{Secrest:2020has, Secrest:2022uvx,Bengaly:2017slg,Siewert:2021,Dam:2022wwh}. In \cite{Dalang:2021ruy}, it was suggested that the redshift evolution of the population of sources may, at least partially, explain the discrepancy. This was further investigated by the authors of \cite{Guandalin:2022tyl}, who also find large variations in the theoretical expectation of the number count dipole in the presence of parameter evolution when using different quasar luminosity function models. The authors of Ref.\,\cite{Dam:2022wwh} reanalyzed the data of \cite{Secrest:2020has, Secrest:2022uvx} and concluded that neither masking nor parameter evolution can fully explain the discrepancy, although the latter is subject to further assumptions. If the dominantly kinematic interpretation of this number count dipole is correct, it should show up in the correlations between the $l$ and $l\pm 1$ multipoles of the number counts, which require high-angular resolution surveys \cite{Dalang:2022gfv,Pant:2018smd}. An observer offset from the center of an ultra-large void was also suggested in \cite{Cai:2022dov} as a solution. This would imply effective large source peculiar velocities as a result of working with a homogeneous and isotropic background. In any case, this problem requires further studies \cite{Aluri:2022hzs}. 

The Hubble tension is somehow more popular \cite{Peebles:2022akh} and has been established for a longer period of time. It is the disagreement between direct measurements of the Hubble constant and inference of $H_0$ from the CMB, if assuming a flat homogeneous and isotropic Universe dominated by cold dark matter and a cosmological constant, the so-called $\Lambda$CDM model. The Hubble constant is inferred from the angle upon which the scale associated to the horizon at the last scattering surface is seen in the CMB, which is extracted from the temperature fluctuations. This results in $H_0 =67.4 \pm 0.5$ km\,s$^{-1}$\,Mpc$^{-1}$ at $68\%$ confidence level \cite{Planck:2018vyg}. In contrast, two of the most competitive local measurements of the Hubble constant come from supernovae type Ia, which requires calibration via the distance ladder and from time-delay cosmography with strongly lensed quasars. Teams performing these experiments reported relatively high $H_0 = 74.03 \pm 1.42$\,km\,s$^{-1}$ \,Mpc$^{-1}$ \cite{Riess:2019cxk, Riess:2021jrx} and $H_0=73.3^{+1.7}_{-1.8}$ km\,s$^{-1}$\,Mpc$^{-1}$ \cite{Wong:2019kwg}, respectively. Combining these two direct measurements leads to a $ 5.3 \sigma$ tension on $H_0$ with the CMB. This has led to a plethora of alternative models, with various levels of complexity and success, as demonstrated by the existence of the $H_0$-olympics \cite{Schoneberg:2021qvd} (see also \cite{DiValentino:2021izs}). 

However, this gravitational lensing estimate of $H_0$ relies on assumptions about the functional form of the mass profile of the lens galaxies and it was pointed out that those choices could result in a bias on $H_0$ if one does not allow for sufficient freedom on the lens model \cite{Kochanek:2020}. Importantly, it should be noted that using only stellar kinematics instead of assumptions about the mass profile of the lensing galaxies to break the so-called mass-sheet degeneracy \cite{Falco:1985}, led to $H_0 = 74.5^{+5.6}_{-6.1}$ km\,s$^{-1}$\,Mpc$^{-1}$ with the seven same strongly lensed systems \citep{Birrer:2020tax}, which is consistent with Planck \cite{Planck:2018vyg}. It was also suggested that the high $H_0$ values from lensing could hint at a dark matter core component in halos \cite{Blum:2020mgu}. Using the stellar kinematics of the SLAC lenses \citep{Bolton2006}, applied to the systems of H0LiCOW with more general lens models, lowers the expectation value of $H_0$ to the Planck value \cite{Birrer:2020tax} but relies on the assumption that the two samples of lenses share similar mass profile properties.

The peculiar velocity of the observer plays the role of a foundational stone for cosmological experiments which work in the CMB frame \cite{Freeman:2005nx,Naselsky_2012,Colin:2019opb,Mohayaee:2021jzi}. It is therefore alarming that some experiments disagree on the peculiar velocity of the observer $\bs{v}_o$. For example, directional dependencies of $2-3\sigma$ level on cosmological parameters extracted from the CMB were reported in \cite{Yeung:2022smn}. A remnant of anisotropies on $H_0$ determined from supernovae type Ia data was reported in \cite{Krishnan:2021dyb} even when working in the CMB frame. The effect of small systematic redshift errors of the order of $10^{-4}$, potentially due to peculiar velocities, was shown to be able to bias $H_0$ obtained from supernovae type Ia to the order of $1$ km s$^{-1}$Mpc$^{-1}$ \cite{Calcino:2016jpu,Davis:2019wet}. The authors of \cite{Cowell:2022ehf} studied anisotropies on $H_0$ paying particular attention to peculiar velocities and found that the monopole of $H_0$ can be biased to the order of $0.30$ km s$^{-1}$ Mpc$^{-1}$, concluding that it is unlikely to fully explain the observed tension.

Galaxy clusters also seem to indicate the presence of $9\%$  anisotropies in $H_0$ as inferred in \cite{Migkas:2021zdo}. Quasars and gamma ray bursts used as standard candles also point toward $2\sigma$ variations in $H_0$ aligned with the CMB dipole \cite{Luongo:2021nqh}. The authors of \cite{Zhai:2022} found $4$ km s$^{-1}$ Mpc$^{-1}$ difference in $H_0$ in opposite hemispheres aligned with the CMB dipole, although such variations are expected. Similar hints of $H_0$ anisotropies from strongly lensed quasars were noticed in \cite{Krishnan:2021jmh}, although these observations are not corrected for the peculiar velocities of the observer, lenses or sources. In particular, the H0LiCOW/TDCOSMO collaboration pointed out a mild $1.8\sigma$ significance for an $H_0$ which decreases with observed lens redshift $z_l'$ \citep{Wong:2019kwg, Millon2020}. Two of the lowest lens redshift systems, which give the highest Hubble constant estimates, turn out to also be well aligned with the CMB dipole, as remarked in \cite{Krishnan:2021jmh}.

In this work, we focus on the determination of the Hubble constant from the time delay of strongly lensed quasars and study the impact of peculiar velocities on this measurement. One may expect that peculiar velocities of the order of $v/c\simeq \mathcal{O}(10^{-3})$ do not affect the $H_0$ measurement beyond $\mathcal{O}(10^{-3})$. However, to our knowledge, there has not been any rigorous study of the accumulation of effects of aberration and Doppler shift on time-delay cosmography, and propagation of biases through the lens model, which may inflate the proportionality constant in front of $v/c$. Our goal is to fill this gap and study if there can be any relation between the Hubble and dipole tensions. 

The paper is structured as follows. In Sec.\,\ref{sec:Singular_Isothermal_Sphere}, we review the basics of time-delay cosmography for a singular isothermal sphere, which allows us to fix notation. In Sec.\,\ref{Sec:Peculiar_velocity_bias}, we detail all effects of peculiar velocities on the observables and also how these propagate through the lens model and to the Hubble constant determination. In Sec.\,\ref{sec:Results}, we apply our findings to the seven lenses of TDCOSMO\footnote{Six lenses come from H0LiCOW \cite{Wong:2019kwg} and one from STRIDES \cite{Shajib:2020}. These seven systems are now analyzed jointly by the TDCOSMO collaboration \cite{Millon2020}.}, compute the bias on the Hubble constant for each lens as a function of the peculiar velocities and discuss our results. Finally, in Sec.\,\ref{sec:Conclusion}, we conclude. We suggest that a busy reader principally interested in the total impact on $H_0$ measurements should review the form of Eqs.\,\eqref{eq:Tij} and \eqref{eq:H0_bias}, then move directly to section \ref{sec:Results}. Throughout the article, bold symbols denote 2 or 3 dimensional vectors, hats indicate unit vectors $||\bs{\hat{n}}||=1$. Sometimes unit vectors in $\mathbb{R}^3$ are expressed in spherical coordinates $\bs{\hat{n}} = (\cos \theta \cos \varphi,\cos \theta \sin\varphi,  \sin \theta)\,\widehat{=}\,(\theta,\varphi) $, where $\theta \in [0,\pi]$ and $\varphi \in [0,2\pi[$ indicate the polar and azimuthal angle, respectively. We note the speed of light $c$ and Newton's constant $G\e{N}$.

\section{Singular Isothermal Sphere}\label{sec:Singular_Isothermal_Sphere}

\begin{figure}
\centering
\includegraphics[width=0.7\textwidth]{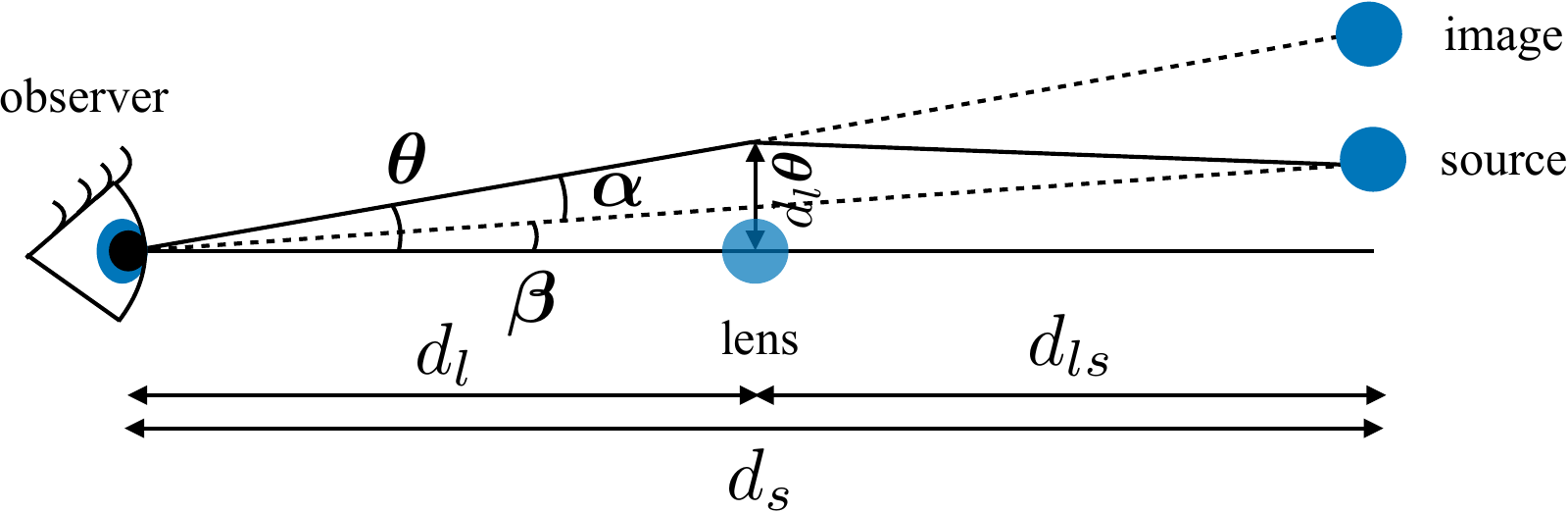}
\caption{We sketch the lensing configuration. The observer, on the left, sees an image at a small angle $\bs{\theta}$ from the optical axis, which connects via a null geodesic the observer to the lens' center of mass. The unobservable angles to the source $\bs{\beta}$ and the deflection angle $\alpha$ are also displayed. The angular diameter distance $d_l$, $d_s$ and $d_{ls}$ at play are displayed in the intuitive Euclidean case where $d_l + d_{ls}=d_s$, although that equality does in general not hold.}
\label{fig:Sketch}
\end{figure}

In this section, we review time-delay cosmography for an isothermal sphere and fix our notation. Derivations may be found in \cite{Schneider} and the reader experienced in lensing time delay formalism can skip to Sec.\,\ref{Sec:Peculiar_velocity_bias}. The cosmic time delay $\Delta t_{ij} \equiv t_i - t_j$ variations in lensed images $i$ and $j$ for a comoving observer, lens and source can be expressed as \cite{Schneider}
\begin{align}
c \Delta t_{ij} = (1+z_l) \frac{d_l d_s}{ d_{ls}} \l[ \hat{\phi}(\bs{\theta}_i, \bs{\beta}) - \hat{\phi}(\bs{\theta_j}, \bs{\beta})\r] 
\,, \label{eq:Time_Delay}
\end{align}
where $z_l$ is the lens redshift, $d_l$, $d_s$ and $d_{ls}$ are angular diameter distances to the lens, to the source and between the lens and the source respectively. A sketch of the lensing configuration is displayed in Fig.\,\ref{fig:Sketch}.  Contrary to Euclidean intuition, $d_l + d_{ls}\neq d_s$, in general. See also \cite{Fleury:2020cal} for a derivation of Eq.\,\eqref{eq:Time_Delay} in arbitrary spacetimes and with arbitrary peculiar velocity configurations.
The dimensionless \textit{Fermat potential} is given by
\begin{align}
\hat{\phi}(\bs{\theta},\bs{\beta}) = \frac{(\bs{\theta}-\bs{\beta})^2}{2} -\psi(\bs{\theta})\,, \label{eq:Fermat_Potential}
\end{align}
where $\bs{\theta}=(\theta_x, \theta_y)$ is a $2$ dimensional vector indicating small observed angles to the images, typically of the order of a few arcsec on the sky, where the origin is the center of mass of the lens, which defines the optical axis. The unobservable $2$ dimensional angle $\bs{\beta} = (\beta_x, \beta_y)$ indicates the source position. This first part of the time delay comes from the geometric difference in the paths followed by photons, emitted simultaneously and deflected by the lens. The \textit{lensing potential} is indicated by $\psi(\bs{\theta})$ and tracks the time delay accumulated by Shapiro time dilation, and requires a lens model to compute. The images form at sky locations which extremize the Fermat potential. In other words, these are solutions of the \textit{lens equation}:
\begin{align}
\bs{\beta} = \bs{\theta}- \bs{\alpha}(\bs{\theta})\,,
\end{align}
where the $2$ dimensional \textit{deflection angle} is $\bs{\alpha}(\bs{\theta}) =(\alpha_x(\bs{\theta}),\alpha_y(\bs{\theta})) = \bs{\nabla} \psi(\bs{\theta})$. Gravitational lenses at cosmological distances have a thickness along the optical axis which can be considered much smaller than the distance between the lens, the source and the observer. In this case, one can make a thin lens approximation to find \cite{Schneider}
\begin{align}
\bs{\alpha}(\bs{\theta}) = \frac{1}{\pi} \int_{\mathbb{R}^2} \dd^2 \bs{\theta'} \kappa(\bs{\theta'}) \frac{\bs{\theta}- \bs{\theta'}}{||\bs{\theta}- \bs{\theta'}||^2}\,, \label{eq:alpha_theta}
\end{align}
where $\kappa(\bs{\theta})$ is the \textit{convergence}, defined as 
\begin{align}
\kappa(\bs{\theta}) \equiv \frac{\Sigma(\bs{\theta})}{\Sigma\e{c}}\,,
\end{align}
where $\Sigma(\bs{\theta})$ is the mass surface density (in kg m$^{-2}$) and the critical surface density is given by
\begin{align}
\Sigma\e{c} \equiv \frac{c^2}{4 \pi G\e{N}} \frac{d_s}{d_{ls} d_l}\,.
\end{align}
Eq.\,\eqref{eq:alpha_theta} expresses that the deflection angle at an angle $\bs{\theta}$ is more affected by the massive regions in the lens plane which are close to $\bs{\theta}$. For a thin lens, the lensing potential can be expressed as
\begin{align}
\psi(\bs{\theta}) & =  \frac{1}{\pi} \int_{\mathbb{R}^2} \dd^2 \bs{\theta'} \kappa(\bs{\theta'}) \log||\bs{\theta}-\bs{\theta'}|| + \hbox{const}\,,
\end{align}
up to an integration constant, which cancels in the time-delay formula. Note that a photon traveling closer (low $||\bs{\theta}-\bs{\theta'}|| \ll 1$) to a region with higher mass density (higher $\kappa(\bs{\theta'})$) will experience more Shapiro time delay (more negative $\psi(\bs{\theta})$) than if it travels far from this region. In the following and throughout the article, we work with a \textit{singular isothermal sphere} (SIS), which can be described by the following mass density
\begin{align}
\rho(r) = \frac{\sigma_v^2}{2 \pi G\e{N} r^2}\,,
\end{align}
in (kg m$^{-3}$) where $r$ is the distance from the center of mass of the lens, $\sigma_v$ is the line-of-sight velocity dispersion, which is assumed to be constant. This lens model is spherically symmetric, singular at $r=0$ and its mass formally extends to infinite radius. Until the end of the 1990s, this was the most popular model for strong lensing time-delay cosmography because all lensing quantities can be derived analytically. The TDCOSMO collaboration has now adopted more sophisticated models to describe the mass profile of the lens galaxy such as the power-law elliptical mass distribution \cite{Barkana1998} and composite models \cite{Suyu2014}, which explicitly includes a baryonic and dark matter component. Nevertheless, we do not expect these more sophisticated lens models to change significantly our results, while the simplicity of the SIS grants us analytic control. The mass surface density can be obtained by integrating along the optical axis, between the source and the observer. This is most easily done in cylindrical coordinates $(r,\varphi, z)$, centered on the lens center of mass. In that case, $\rho(r) = \rho(d_l\theta,0,l)$ with $\theta=||\bs{\theta}||$ and we get
\begin{align}
\Sigma(\bs{\theta}) = \int_{l_s}^{l_o} \dd l \rho(d_l\theta,0,l) = \frac{\sigma_v^2}{2 \pi G\e{N} d_l \theta} \hbox{Arccot}\l(\frac{d_l \theta}{l}\Big|^{l_o}_{l_s}\r)\,.
\end{align}
Taking the limit of far away source and observer, compared to the impact parameter $|l_o|,|l_s| \gg d_l \theta$, one finds
\begin{align}
\Sigma(\theta) = \frac{\sigma_v^2}{ G\e{N} d_l \theta}\,.
\end{align}
Making use of axial symmetry, (i.e.\,$\kappa(\bs{\theta}) = \kappa(\theta)$), one finds, using Eq.\,\eqref{eq:alpha_theta}, that $\bs{\alpha}(\bs{\theta}) = \alpha(\theta) \bs{\theta}/\theta$ with
\begin{align}
\alpha(\theta) = \frac{2}{\theta} \int_0^\theta \dd \theta' \theta' \kappa(\theta')\,.
\end{align}
For an SIS, this integral reduces to a constant deflection angle
\begin{align}
\alpha(\theta) = \frac{4\pi \sigma_v^2}{c^2} \frac{d_{ls}}{d_s}\equiv \alpha_0\,.\label{eq:alpha_0}
\end{align}
This implies that the source angle $\bs{\beta}$ for an SIS can be reconstructed from only one image $\bs{\theta_i}$,
\begin{align}
\bs{\beta} = \bs{\theta_i}\l(1-\frac{\alpha_0}{\theta_i}\r)\,.
\end{align}
This can also be read as a quadratic equation for $\bs{\theta_i}$, which gives at most 2 images\footnote{There is also a third image at $\theta_i=0$, which is infinitely demagnified.}. In practice, external shear or deviations from spherical symmetry of the lens can lead to the formation of $N\e{images}>2$ images. This implies that if one attempts to reconstruct $\bs{\beta}$ for these systems, one may get slightly different results for each image, which affect the determination of the Hubble constant. Therefore, for practical purposes, one rather estimates a source angle for each image $\bs{\beta}= \bs{\beta}(\bs{\theta_i})$. Similarly, the lensing potential for an axially symmetric thin lens can be expressed as
\begin{align}
\psi(\bs{\theta}) = 2 \int_0^\theta \dd \theta' \theta' \kappa(\theta') \log(\theta/\theta') +\hbox{const.}\,.
\end{align}
which reduces to 
\begin{align}
\psi(\theta) = \alpha_0 \theta\,, \label{eq:Psi_SIS}
\end{align}
for an SIS. One can recognize the primitive of $\alpha$ (Eq.\,\eqref{eq:alpha_0}), where the integration constant has been set to zero. In practice, the angular diameter distances are not measured directly but can be inferred from the lens and source redshifts $z_l$ and $z_s$, by assuming a cosmological model.\footnote{Note that if one would measure the angular diameter distances directly, one could check that the time-delay formula holds for arbitrary peculiar velocity configurations \cite{Fleury:2020cal}.} Throughout the article, we assume a flat $\Lambda$CDM model with $\Omega_{m0}=0.3$ and $H_0 =70$ km s$^{-1}$ Mpc$^{-1}$. Of course, the determination of $H_0$ from observables does not require an assumption on $H_0$ but the relative bias, as we will find in Sec.\,\ref{Sec:Peculiar_velocity_bias}, does depend on $H_0$. The angular diameter distances can be expressed as 
\begin{align}
d_l & = d_l[z_l] = \frac{c }{H_0(1+z_l)} \chi[z_l]\,, \label{eq:dl0}\\
d_s & = d_s[z_s] = \frac{c}{H_0(1+z_s)}\chi[z_s]\,, \label{eq:ds0} \\
d_{ls} & =d_{l}[z_l,z_s] = \frac{c}{H_0(1+z_s)} \chi[z_l,z_s]\label{eq:dls0}\,,
\end{align}
where $H_0$ is the present-day Hubble constant, $\chi[z_1,z_2]$ is the dimensionless integral
\begin{align}
\chi[z_1,z_2] \equiv \int_{z_1}^{z_2} \frac{\dd z }{E(z)}\,, \label{eq:chi}
\end{align}
with $H[z]= H_0 E(z) \equiv H_0 \sqrt{\Omega_{m0}(1+z)^3 + (1-\Omega_{m0})}$ and $\chi[z]=\chi[0,z]$, which should not be confused with comoving distances. One can solve Eq.\,\eqref{eq:Time_Delay} for $H_0$ to get
\begin{align}
H_0 = \frac{\chi[z_l]\chi[z_s]}{\chi[z_l,z_s]} \frac{\hat{\phi}(\bs{\theta_i},\bs{\beta}(\bs{\theta_i})) - \hat{\phi}(\bs{\theta_j},\bs{\beta}(\bs{\theta_j}))}{\Delta t_{ij}}\,.
\end{align}
The present-day Hubble constant is expressed in terms of the lens and source redshifts $z_l$, $z_s$, the time delay $\Delta t_{ij}$, the images $\bs{\theta}_i$, $i, j \in [1,\dots, N\e{images}]$ and the velocity dispersion of the lens $\sigma_v$. Nearly all of these observables are directly affected by peculiar velocities to some extent; some are also indirectly affected through the lens model, and we detail how in the next section.

\section{Peculiar velocity bias}\label{Sec:Peculiar_velocity_bias}
The previous section outlined how one may relate the present-day Hubble rate to time-delayed images of a lensed source, assuming a comoving observer, lens and source. In this section, we relax this assumption and compute the bias that the nonrelativistic peculiar velocities of the observer $\bs{v}_o$, the lens $\bs{v}_l$ and the source $\bs{v}_s$ generate on $H_0$ to linear order in $v/c \ll 1$, where $v$ indicates any of the three peculiar velocities. In particular, we detail the computation of the biases, which are quite straightforward for time delays, redshift and angular diameter distances as a function of redshift. On the other hand, the effect of aberration of angles turns out to be quite subtle, especially to infer reconstructed quantities like the source angle or the lensing potential. Time-pressured readers may directly skip to Eq.\,\eqref{eq:Tij} and \eqref{eq:H0_bias}, which constitute the main results of this section. We denote the quantities that are observed with a prime, while the quantities that comoving (virtual) observers\footnote{It turns out that it is extremely unlikely to be a comoving observer. In particular, in a Universe with structures such as galaxies and filaments, the probability for a massive observer to be comoving is zero. Observers on Earth are certainly not.} would measure are left without a prime.

\subsection{Time dilation} \label{sec:Time_Dilation}
The motion of the observer induces a special relativistic time dilation, which prevents them from measuring cosmic time directly. However, this effect is second order in the velocity of the observer and we neglect it:
\begin{align}
\Delta t_{ij}' = \frac{\Delta t_{ij}}{\sqrt{1-\bs{v}_o^2/c^2}}= \Delta t_{ij}[1+\mathcal{O}(\bs{v}_o^2/c^2)]\,. \label{eq:Time_Dilation}
\end{align}
The velocity of the source does not affect the observed time delay because one observes the time delay between 
flux variations of the quasar that have been emitted simultaneously.

\subsection{Redshifts}\label{sec:Redshifts}
The motion of the observer, lens and source affect the lens and source \textit{observed} redshifts with respect to background (cosmological) redshifts through Doppler shift. The observed redshifts $z_l'$, $z_s'$ relate to cosmological (or background) redshift $z_l$, $z_s$ in the following way
\begin{align}
(1+z_l) & = (1+z_l')\l(1+Z_L \frac{v_o}{c} \r)\,, \label{eq:zl}\\
(1+z_s) & = (1+z_s')\l(1+Z_S \frac{v_o}{c} \r) \label{eq:zs}
\end{align}
with
\begin{align}
Z_L & =\frac{\bs{\hat{n}'} \cdot (\bs{v}_o - \bs{v}_l)}{v_o} \,,\\
Z_S & = \frac{\bs{\hat{n}'} \cdot (\bs{v}_o - \bs{v}_s)}{v_o}\,.
\end{align}
This apparent expansion in $v_o/c$ in Eqs.\,\eqref{eq:dl_1}-\eqref{eq:dls_1} is practical for bookkeeping, but one should keep in mind that it really is a simultaneous expansion in $v_o/c$, $v_l/c$ and $v_s/c$. This affects the time delay (Eq.\,\eqref{eq:Time_Delay}) via the lens redshift $z_l$ and via the background angular diameter distances, which can be computed from the redshift information. Throughout this work, we denote biases on a quantity by a corresponding capital letter, which carries the same units (e.g.\,$Z_S$ is the bias generated by peculiar velocities on $z_s$).

\subsection{Angular diameter distances}\label{sec:Angular_Diameter_Distances}
One can compute the background angular diameter distances from the observed redshift by assuming a cosmological model, provided one corrects for the peculiar motion of the emitter and receiver. By background angular diameter distance, we mean the distance that would be inferred by a comoving observer that would measure the subtended angle on the sky of a standard ruler. For example, the background angular diameter distance to the source can be expressed as a function of observed redshift $z_s'$
\begin{align}
d_s & = \frac{c}{1+z_s(z_s')} \int_{0}^{z_s(z_s')} \frac{\dd z}{H(z)} \\
& = \frac{1}{1+z_s(z_s')} \int_{0}^{z_s'+(1+z_s')\bs{\hat{n}'}\cdot  (\bs{v}_o-\bs{v}_s)} \frac{\dd z}{H(z)} \\%\\
%& =\frac{c}{1+z_s'}\l( 1- \frac{\bs{\hat{n}'} \cdot (\bs{v}_o - \bs{v}_s)}{c}\r) \l( \int_{0}^{z_s'} \frac{\dd z}{H(z)} + \int_{z_s'}^{z_s'+(1+z_s')\bs{\hat{n}'}\cdot (\bs{v}_o-\bs{v}_s)} \frac{\dd z}{H(z)}\r) \\
& \simeq d[z_s'] +  \frac{\bs{\hat{n}'}\cdot (\bs{v}_o-\bs{v}_s ) }{c} \l( \frac{c}{H(z_s')} - d[z_s']\r)\,,
\end{align}
where $d[z_s']$ indicates the naive background angular diameter distance as a function of observed redshift, as given in Eq.\,\eqref{eq:dsp}. Therefore, one can compute the background angular diameter distances $d_l$, $d_s$ and $d_{ls}$ as follows
\begin{align}
d_l &  = d[z_l'] + D_L \frac{v_o}{c}\,,\label{eq:dl_1}\\
d_s &  = d[z_l'] + D_S \frac{v_o}{c}\,, \label{eq:ds_1}\\
d_{ls}&  = d[z_l',z_s'] + D_{LS} \frac{v_o}{c}\,,\label{eq:dls_1}
\end{align}
with
\begin{align}
d[z_l'] & = \frac{c}{H_0 (1+z_l')} \chi[z_l']\,, \label{eq:dlp}\\
d[z_s'] & = \frac{c}{H_0 (1+z_s')} \chi[z_s']\,, \label{eq:dsp}\\
d[z_l', z_s'] & = \frac{c}{H_0 (1+z_s')} \chi[z_s',z_s']\,, \label{eq:dlsp}
\end{align}
where the function $\chi$ was defined explicitly in Eq.\,\eqref{eq:chi} and where the lens and source peculiar velocities are included in the corrections
\begin{align}
D_L & =  \frac{\bs{\hat{n}'}\cdot(\bs{v}_o-\bs{v}_l)}{v_o} \l( \frac{c}{H[z_l']}  - d[z_l']\r)\,, \label{eq:DL} \\
D_S & =   \frac{\bs{\hat{n}'}\cdot(\bs{v}_o-\bs{v}_s)}{v_o} \l( \frac{c}{H[z_s']} - d[z_s'] \r)\,,\label{eq:DS} \\
D_{LS} & =\frac{\bs{\hat{n}'}\cdot(\bs{v}_o-\bs{v}_s)}{v_o} \l(  \frac{c}{H[z_s']}  -d[z_l',z_s'] \r) - \frac{c }{H[z_l']} \frac{1+z_l'}{1+z_s'} \frac{\bs{\hat{n}'}\cdot(\bs{v}_o-\bs{v}_l)}{v_o} \,. \label{eq:DLS}
\end{align}
The distances $d[z_l']$, $d[z_s']$ and $d[z_l',z_s']$ are the naive background angular diameter distances, which can be computed from Eqs.\,\eqref{eq:dlp}-\eqref{eq:dlsp}. Note that it is only the background angular diameter distances as a function of observed redshifts which are biased in this way. As encountered with Eq.\,\eqref{eq:zl}-\eqref{zs}, we remind the reader that this apparent expansion in $v_o/c$ really is a simultaneous expansion in $v_o/c$, $v_l/c$ and $v_s/c$. The projection of the lens and source peculiar velocities along the line of sight are unknown and difficult to measure. We shall vary $v_l^\parallel \equiv \bs{\hat{n}'}\cdot \bs{v}_l $ and  $v_s^\parallel \equiv \bs{\hat{n}'}\cdot \bs{v}_s $ to quantify their impact. 

\subsection{Aberration of angles} \label{sec:Aberration_of_angles}

In this technical subsection, we give explicit expressions to compute the bias generated by peculiar velocities on the measured angles to the images, the Einstein angle and on the inferred source angle. The main results are the biases on these three angles, which can be found in Eqs.\,\eqref{eq:T}, \eqref{eq:sigma_v} and \eqref{eq:betapp_result}. 

Observed angles on the sky are affected by the peculiar velocity of the observer. It appears simpler to compute the effect of aberration in a frame in which the $\bs{\hat{z}}$ axis coincides with the direction of the peculiar velocity of the observer $\bs{v}_o$. In this special case, only the polar angle $\theta$ is affected by the boost, while the azimuthal angle $\varphi$ is left unaffected to first order
\begin{align}
\theta' & = \theta -\sin (\theta) \frac{v_o}{c}\,, \label{eq:Angle_Transformation}\\
\varphi' & = \varphi\,.
\end{align}
Note that to first order in $v_o/c$, one can easily invert the system
\begin{align}
\theta &= \theta' + \sin (\theta') \frac{v_o}{c}\,, \label{eq:Angle_Transformation_2}\\
\varphi & =\varphi'\,.\label{eq:Angle_Transformation_phi}
\end{align} 
While this is convenient from a calculational point of view, it requires to translate the observations into that coordinate system, which we call the \textit{calculation} coordinate system. To this end, we also introduce an \textit{observation} coordinate system, which carries tildes, which are 2-dimensional angles on the sky in the neighborhood of the lens' center of mass, which corresponds to the origin that points toward $\bs{\hat{n}'}$. The $\bs{\t{\theta}'_y}$ vector is the projection of the North pole (J2000) in the plane orthogonal to $\bs{\hat{n}'}$, while $\bs{\t{\theta}'_x}$ points East. This is the coordinate system in which strong lensing observations are made. Images are couples $\bs{\t{\theta}_i'}= (\t{\theta}'_{ix},\t{\theta}'_{iy})$ in that coordinate system. There is one such coordinate system for observers with peculiar velocity $\bs{v}_o$, which carries primes on top of tildes and one for comoving observers (that have $\bs{v}_o =0$), which is free of primes.
\begin{figure}
    \centering
    \includegraphics[width=0.40\textwidth]{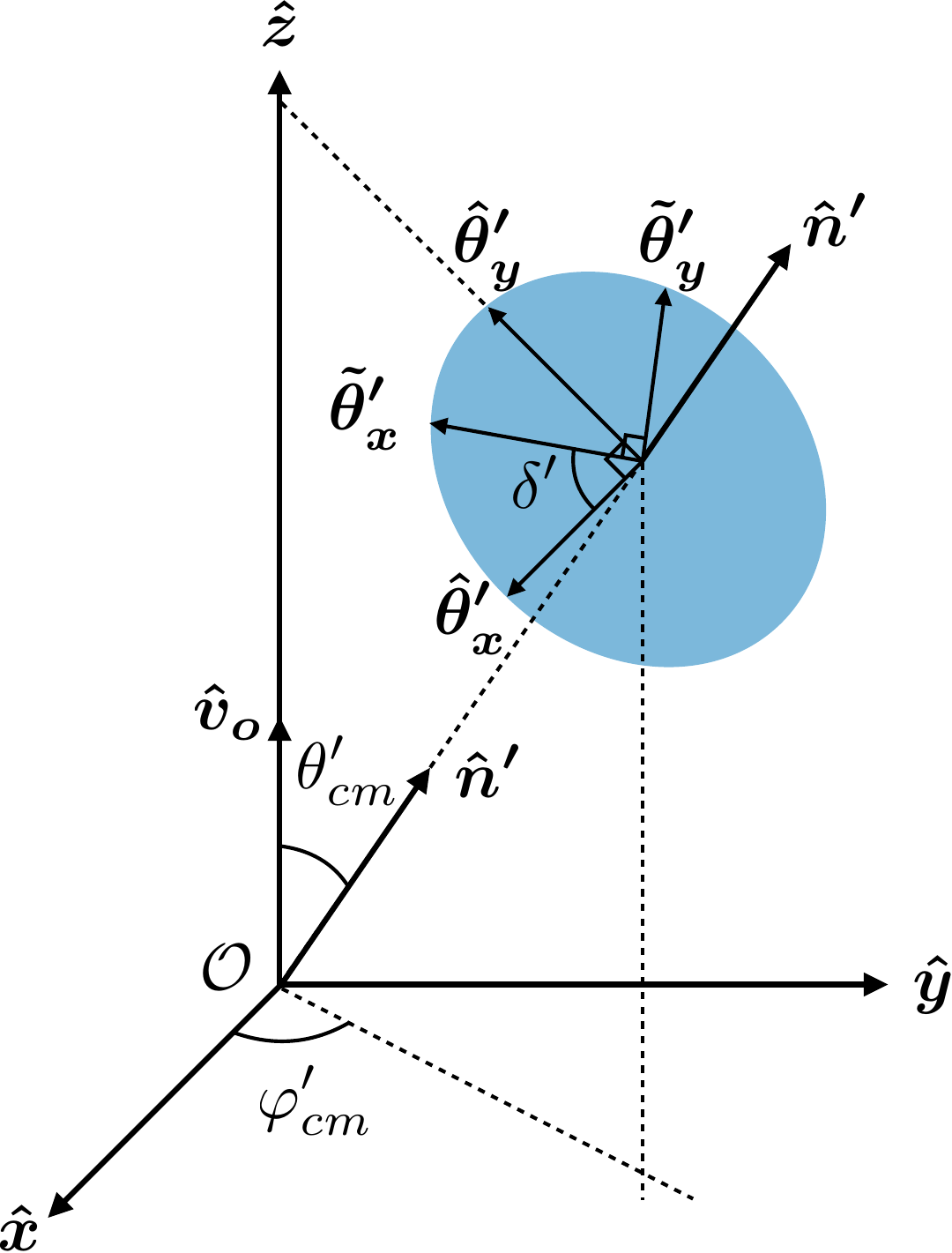}
    \caption{We plot here the coordinate systems involved. Both observers sit at the origin $\mathcal{O}$. One observer is at rest in this coordinate system and would observe comoving quantities, which have no primes. The observer moving with peculiar velocity $\bs{v}_o$ which is aligned with $\bs{\hat{z}}$ works in the observation coordinate system, spanned by the two vectors $\{\bs{\t{\theta}_x'}, \bs{\t{\theta}_y'}\}$ which are denoted with primes. The vector $\bs{\t{\theta}_y'}$ is the projection of the Earth's North pole direction in the plane orthogonal to $\bs{\hat{n}'}$, while $\bs{\t{\theta}'_x}$ points East. The moving observer sees the lensed system center of mass in the direction $\bs{\hat{n}'} = (\theta_{cm}', \varphi_{cm}')$. The more convenient basis is the hatted one, which is spanned by $\{ \bs{\hat{\theta}_x'},\bs{\hat{\theta}_y'}  \}$. This convenient coordinate system is such that $\bs{\hat{\theta}_x'}$ belongs to the plane orthogonal to $\bs{\hat{z}}$. As such, it is unaffected by the boost. The angle $\delta'$ relates the two basis such that $\cos \delta' = \bs{\hat{\theta}'_x} \cdot \bs{\tilde{\theta}'_x}$. }
    \label{fig:Coordinate_System}
\end{figure}

\subsubsection{Distortion of the images}

\begin{figure}
    \centering
    \includegraphics[width=0.80\textwidth]{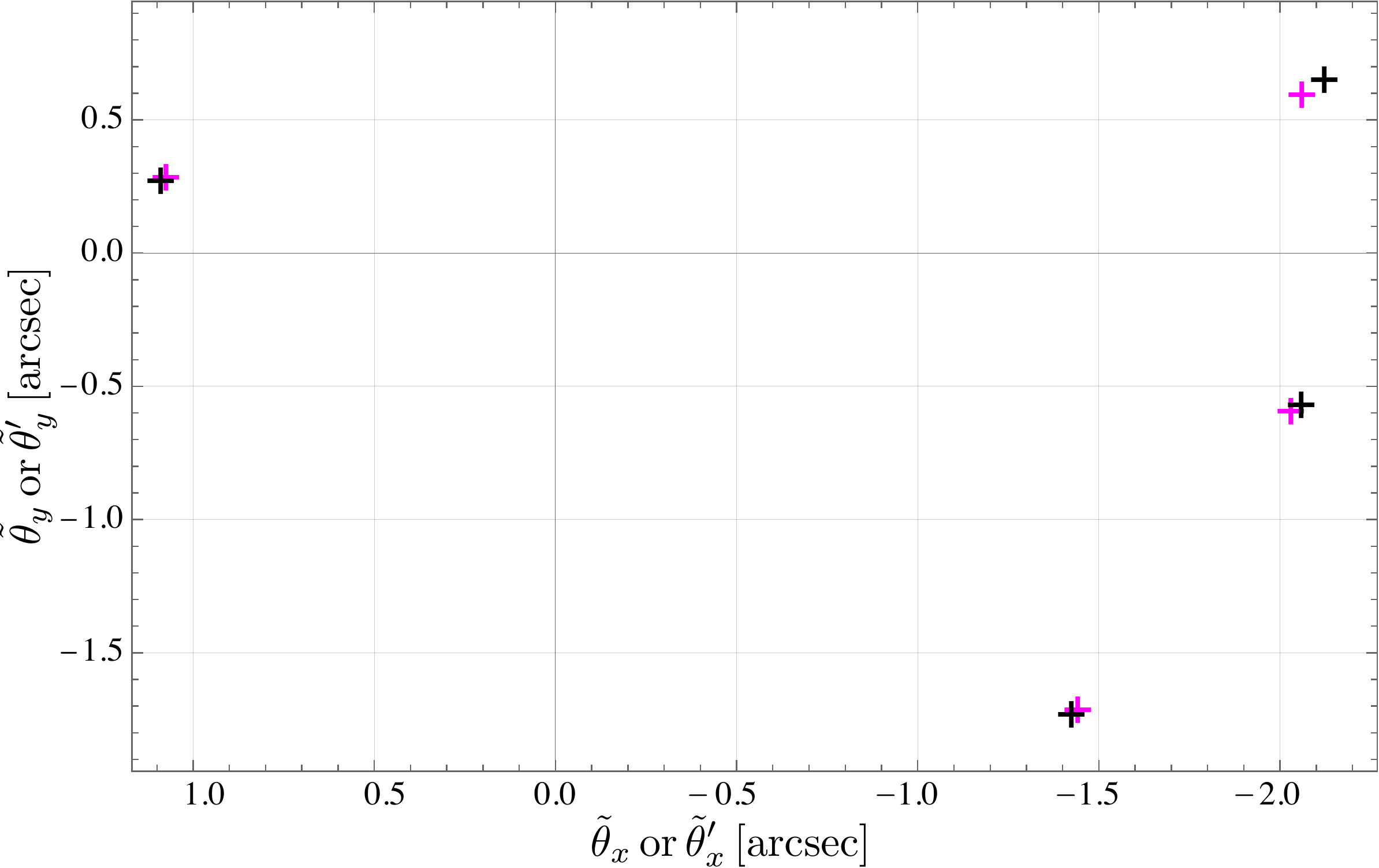}
    \caption{We plot the 4 images $\bs{\t{\theta}'_i}$ of RXJ1131-1231 (in pink) and the corresponding images $\bs{\t{\theta}_i}$ (in black) that would be seen by a comoving observer if $\bs{v}_o = 40 \,\bs{v}\e{dip}$. Each image is displaced by $\bs{\Theta}_i v_o/c$, as should be clear from Eq.\,\eqref{eq:T}. The origin on this plot corresponds to the directions of $\bs{\hat{n}'}$ and $\bs{\hat{n}}$ in the appropriate cases. Note that the $\bs{\t{\theta}_x}$ and the $\bs{\t{\theta}'_x}$ axis point in different directions which are captured by $\delta$ and $\delta'$, as in Eqs.\eqref{eq:thetap_i}-\eqref{eq:varphi_i}.} 
    \label{fig:Image_distortions}
\end{figure}

Each image appears to a boosted observer with polar and azimuthal angles $\{ \theta_i',\varphi_i'\}$. These can be computed, given an observed center of mass lens $\bs{\hat{n}'} = (\theta_{cm}', \varphi_{cm}')$, a rotation angle $\delta'$, which can be computed for a given $\bs{\hat{n}'}$ following App.\,\ref{app:Rotation_angle} and image coordinates $\bs{\t{\theta}'_i}$
\begin{align}
\theta_i' = \theta_{cm}' - \hat{\theta}_{iy}' =  \theta_{cm}' - \l( \t{\theta}_{ix}' \sin \delta' + \t{\theta}_{iy}' \cos\delta' \r)\,,\label{eq:thetap_i} \\
\varphi_i' = \varphi_{cm}' - \hat{\theta}_{ix}' = \varphi_{cm}' - \l(\t{\theta}_{ix}' \cos \delta' - \t{\theta}_{iy}' \sin \delta' \r)\,.
\end{align}
Applying Eq.\,\eqref{eq:Angle_Transformation_2}-\eqref{eq:Angle_Transformation_phi} to infer $\bs{\hat{n}}$ and $\{ \theta_i, \varphi_i\}$, one can solve the following system for $\bs{\t{\theta}_i}$
\begin{align}
\theta_i &= \theta_{cm} - \l( \t{\theta}_{ix} \sin \delta + \t{\theta}_{iy} \cos\delta \r)\,, \label{eq:theta_i}\\
\varphi_i & = \varphi_{cm} - \l(\t{\theta}_{ix} \cos \delta - \t{\theta}_{iy} \sin \delta \r)\,, \label{eq:varphi_i}
\end{align}
where in particular $\delta \neq \delta'$, in general (see App.\,\ref{app:Rotation_angle}). One then defines the bias $\bs{\T}_i = (\T_{ix},\T_{iy})$ on image $i$ implicitly as
\begin{align}
\bs{\t{\theta}}_i = \bs{\t{\theta}'_i} + \bs{\T_i} \frac{v_o}{c}\,. \label{eq:T}
\end{align}
This equation can be used to compute $\bs{\T_i}$ from the observed images $\bs{\t{\theta}_i'}$ together with the solutions $\bs{\t{\theta}_i}$ of Eqs.\,\eqref{eq:theta_i}-\eqref{eq:varphi_i}, $v_o$ and rotation angles $\delta$, $\delta'$ given in App.\,\ref{app:Rotation_angle}. Note that this bias is independent of the peculiar velocity of the lens and source. The images are affected in slightly different ways, due to their different sky positions relative to $\bs{\hat{v}}_o$. This cannot be captured by an image-independent translation for one lens, as can be seen from Fig.\,\ref{fig:Image_distortions}, where we plot the displaced images for the system RXJ1131-1231 for the exaggerated case $\bs{v}_o=40 \bs{v}\e{dip}$. This is why we rather speak of image distortion, rather than translation.

\subsubsection{The velocity dispersion from the Einstein angle}

The central velocity dispersion of the lens galaxy traces its total mass and can be either measured directly from spectroscopic observation or deduced from the Einstein radius with some assumptions about the mass profile of the lens. In the former case, this quantity can in principle be measured independently of peculiar velocities, since these would only affect the position of the spectral lines while leaving their width unchanged. The velocity dispersion inferred from the spectral lines' width would therefore be unaffected. However, velocity dispersions obtained with this technique are limited to a precision of $\sim 10$ \%, which is not sufficient to precisely constrain the mass profile of the lens galaxies. In fact, most of the constraints on the mass profile in recent time-delay cosmography analysis come from the lensing observables, including the Einstein radius. Since the Einstein radius is affected by the aberration on the measured angle described in the previous section, this error propagates to the mass profile. In this subsection, we use the central velocity dispersion of the lens, $\sigma_v$ as a proxy to quantify the error on the mass profile due to the aberration on the measured Einstein angle. The Einstein angle can be related to $\sigma_v$ from the following relation \cite{Schneider}
\begin{align}
\theta_E = \frac{4\pi \sigma_v^2}{c^2} \frac{d_{ls}}{d_s}
\end{align}
for an SIS. This angle corresponds to the angle under which an observer perfectly aligned with the lens and a pointlike source would see an Einstein ring. Note that it matches $\alpha_0$, defined in Eq.\,\eqref{eq:alpha_0}. For simplicity, we assume that one measures the Einstein angle in a plane which is spanned by $\bs{\hat{v}}_o$ and $\bs{\hat{n}'}$ (that is, in direction $\bs{\hat{\theta}'_y}$ (see Fig.\,\ref{fig:Coordinate_System})). In this case, the aberration of the Einstein ring is maximal. One finds
\begin{align}
\theta_E & = \theta_E' + \frac{v_o}{c} ( \sin(\theta_{cm}' + \theta_E') - \sin(\theta_{cm}')) \\
& = \theta_E' + \frac{v_o}{c} \cos(\theta_{cm}') \theta_E'\,.
\end{align}
In this case, biased measurements of $z_l'$, $z_s'$ and $\theta_E'$ of $z_l$, $z_s$ and $\theta_E$ induce a bias on the inference $\sigma_v'$ of $\sigma_v$. It can be estimated to first order in the peculiar velocities by
\begin{align}
\sigma_v = \sigma_v' + S_v \frac{v_o}{c} \,,\label{eq:sigma_v}
\end{align}
with
\begin{align}
\sigma_v' & = \sqrt{\frac{d[z_s'] }{d[z_l',z_s']} \frac{\theta_E'}{4\pi}} c\,,\\
S_v & = \frac{\sigma_v'}{ 2 d[z_l',z_s'] d[z_s']}  \big[d[z_l',z_s']D_S  - d[z_s'] D_{LS}  + d[z_l',z_s'] d[z_s'] \cos \theta_{cm}' \big]\,. \label{eq:S_v}
\end{align}
Here the distances $d[z_s']$, $d[z_l',z_s']$ and their related biases $D_S$ and $D_{LS}$ can be computed using Eqs.\,\eqref{eq:dlp}-\eqref{eq:DLS}, which depend on the source, lens and observer's peculiar velocities. Note that we use capital letters to denote biases, not the angular diameter distances themselves. This is rather an overestimation of the bias on $\sigma_v$, if estimated from the observed Einstein angle. This is because angles, including the Einstein angle, are unaffected\footnote{This is the reason why the intermediate coordinate system spanned by $\{\bs{\hat{\theta}_x'}, \bs{\hat{\theta}_y'}\}$ was introduced.} in the direction $\bs{\hat{\theta}_x'}$. It turns out that the bias $S_v$ on $\sigma_v$ increases the bias on $H_0$ generated by peculiar velocities. In the quantitative analysis presented in Sec.\,\ref{sec:Results}, we shall also study what happens if one measures $\sigma_v$ independently (setting $S_v=0$), by direct peculiar velocity dispersion measurements in redshift space. This would also correspond to the situation in which the Einstein angle is measured in the direction $\bs{\hat{\theta}_x'}$. In practice, one can measure the azimuthally averaged Einstein radius. A perfect circle Einstein ring seen by a comoving observer would be unaffected in the direction $\bs{\hat{\theta}_x'}$ and maximally affected in the direction $\bs{\hat{\theta}_y'}$. Whether the enclosed area of the deformed circle is larger or smaller depends on the sign of $\cos(\theta_{cm}')$. We expect the practical case to lie somewhat in between these two situations.

\subsubsection{The reconstructed source angle}

Reconstructing the source angle $\bs{\t{\beta}}$ is subtle. This is because it is a quantity which is \textit{inferred}, as opposed to \textit{observed}, from biased observations like $\bs{\t{\theta}'}$, $z_l'$ and $z_s'$ and that it appears directly in the time-delay formula (Eq.\,\eqref{eq:Time_Delay}). Here, we write a tilde, to remind the reader that it is a two-dimensional angle in the observation coordinate system. The reconstruction of $\bs{\t{\beta}}$ consists of two steps. The first one consists in estimating the angle $\bs{\t{\beta}''}$ directly from the observed quantities $\bs{\t{\theta}'}$, $z_l'$, $z_s'$. The angle $\bs{\t{\beta}'}$ to the source which would be observed in absence of the lens can also be computed from these observables and knowledge of the peculiar velocities. In the second step, one can reconstruct the angle to the source $\bs{\t{\beta}}$ that a \textit{comoving} observer would observe in absence of the lens. We carry on with the first step. The lens equation for a singular isothermal sphere and comoving observer, source and lens reads
\begin{align}
\bs{\t{\beta}} = \bs{\t{\theta}}\l(1 -  \frac{\alpha_0}{||\bs{\t{\theta}}||}\r)\,. \label{eq:betat}
\end{align} 
This equation allows, through the observation of images $\bs{\t{\theta}}_i$ and an estimate of $\alpha_0$ to reconstruct $\bs{\t{\beta}}$. However, all of these quantities are affected by the boost and so is the reconstruction of $\bs{\t{\beta}}$. By measuring $\theta_E'$, $z_l'$ and $z_s'$, one estimates $\alpha_0'$, which is related to a comoving deflection angle $\alpha_0$ by
\begin{align}
\alpha_0 = \alpha_0' + A_0 \frac{v_o}{c}\,,
\end{align}
with
\begin{align}
\alpha_0' & = \frac{4\pi (\sigma'_v)^2}{c^2 } \frac{d[z_l',z_s']}{d[z_s']}\,,\\
A_0& = \frac{4 \pi \sigma_v'}{c^2 d^2[z_s']} \big( 2 d[z_l', z_s'] d[z_s'] S_v - d [z_l', z_s']  D_S \sigma_v' + D_{LS} d[z_s'] \sigma_v' \big)\,. \label{eq:A_0}
\end{align}
Here the distances $d[z_s']$, $d[z_l',z_s']$, their biases $D_S$, $D_{LS}$ and $S_v$ can be calculated directly from the observables, using Eqs.\,\eqref{eq:dlp}-\eqref{eq:DS} and \eqref{eq:S_v}. The deflection angle is therefore biased by the distance biases and the bias on the veloctiy dispersion. The inference of $\bs{\t{\beta}}'$ as should be made by an observer with peculiar velocity $v_o$ is biased because of the bias in all images $\bs{\t{\theta}_i}'$, redshift of the lens and source and because of the bias in $\alpha_0$. There is only one true source angle $\bs{\t{\beta}}$. However, since we use an isothermal sphere, which has only 2 images; for systems which have 3 or 4 images, the $\bs{\t{\beta}}$ inferred via Eq.\,\eqref{eq:betat} may give different results depending on which image is used. This turns out to impact significantly the determination of the Hubble constant. Therefore, we compute $\bs{\t{\beta}'_i}$ and its corresponding bias $\bs{B'_i} $ for each image. We get
\begin{align}
\bs{\t{\beta}'_i} & = \bs{\t{\beta}''_i} + \bs{B'_i} \frac{v_o}{c} \,,
\end{align}
with
\begin{align}
\bs{\t{\beta}''_i} & = \bs{\t{\theta}'_i} \l(1 -  \frac{ \alpha_0'}{||\bs{\t{\theta}_{i}'}||} \r)\,, \label{eq:betapp}
\end{align}
and
\begin{align}
B_{ix}' & = \T_{ix} + \frac{1}{||\bs{\t{\theta}_{i}'}||^3}  \Big( \alpha_0' \t{\theta}_{iy}' (\T_{iy} \t{\theta}_{ix}' - \T_{ix} \t{\theta}_{iy}' ) - A_0 \t{\theta}_{ix}' ||\bs{\t{\theta}_i'}||^2 \Big )\,, \label{eq:Bpx}\\
B_{iy}' & = \T_{iy} + \frac{1}{||\bs{\t{\theta}_{i}'}||^3}  \Big( \alpha_0' \t{\theta}_{ix}' (\T_{ix} \t{\theta}_{iy}' - \T_{iy} \t{\theta}_{ix}' ) - A_0 \t{\theta}_{iy}' ||\bs{\t{\theta}_i'}||^2 \Big )\,, \label{eq:Bpy}
\end{align}
where $\bs{\T_i}$ and $A_0$ were defined in Eq.\,\eqref{eq:T} and \eqref{eq:A_0}. Those can be computed directly from the observables. A moving observer makes a biased inference $\bs{\t{\beta}}''$ of $\bs{\t{\beta}}'$, which differs from image to image. We wish to express this source angle on the sky $\bs{\beta'}=(\theta_{\beta}',\varphi_\beta' )$ for a comoving observer, which would rather observe $\bs{\beta}=(\theta_\beta, \varphi_\beta)$, given by
\begin{align}
\theta_\beta & = \theta_\beta' + \frac{v_o}{c} \sin \theta_\beta'\,, \label{eq:Comoving_Theta_Beta}\\
\varphi_\beta & = \varphi_\beta'\,,
\end{align}
where the right hand side can be computed directly by the measured quantities
\begin{align}
\theta_\beta' & = \theta_{cm}' - (\t{\beta}'_x \sin \delta' + \t{\beta}_y' \cos \delta' )\,,\\
\varphi_\beta' & = \varphi_{cm}' - (\t{\beta}_x' \cos \delta' - \t{\beta}_y' \sin \delta') \,,
\end{align}
together with the rotation angle $\delta'$, which can be computed for a given direction following App.\,\ref{app:Rotation_angle}. 
Once the left hand side of Eq.\,\eqref{eq:Comoving_Theta_Beta} is determined, one can infer $\bs{\t{\beta}}$ that would be inferred by a comoving observer by solving the following equations for $\bs{\t{\beta}}$
\begin{align}
\theta_\beta & = \theta_{cm} - (\t{\beta}_x \sin \delta  + \t{\beta}_y \cos\delta )\,,\\
\varphi_\beta & = \varphi_{cm} - (\t{\beta}_x  \cos \delta - \t{\beta}_y \sin \delta )\,,
\end{align}
where $\delta \neq \delta'$ can also be computed following App.\,\ref{app:Rotation_angle}. The solutions can be expressed as
\begin{align}
\bs{\t{\beta}_i} & = \bs{\t{\beta}'_i} + \bs{B_i} \frac{v_o}{c}\,, \label{eq:B}
\end{align}
where the image index $i$ was reintroduced and which defines implicitly the bias $\bs{B_i}$. Note that in general, $\bs{B_i}\neq \bs{B'_i}$. In this way,
\begin{align}
\bs{\t{\beta}_i} & = \bs{\t{\beta}''_i} + \bs{B''_i} \frac{v_o}{c}\,, \label{eq:betapp_result}\\
\bs{B''_i} & \equiv \bs{B'_i}+ \bs{B_i}\,, \label{eq:Bpp}
\end{align}
where $\bs{B'_i}$ was defined in Eqs.\,\eqref{eq:Bpx}-\eqref{eq:Bpy} and $\bs{B}_i$ was defined implicitly in Eq.\,\eqref{eq:B}. Those can be computed directly from the observables. In this sense, one pays twice the price in neglecting peculiar velocities in the determination of $\bs{\t{\beta}}$. That is because it is a quantity which is inferred from biased quantities like $\bs{\t{\theta}_i'}$, $z_l'$ and $z_s'$. One first needs to reconstruct the angle to the source $\bs{\t{\beta}'}$ that the \textit{moving} observer would see in absence of the lens. Only then, one can compute the angle to the source $\bs{\t{\beta}}$ that would be seen by a \textit{comoving} observer. Eqs.\,\eqref{eq:betapp_result} and \eqref{eq:Bpp} are the final results of this section, which we use for the remainder of this work. The source angle is biased by the source, lens and observer's peculiar velocities. 

\subsection{The lensing potential}\label{sec:Lensing_Potential}
The lensing potential for an isothermal sphere reads (see Eq.\,\eqref{eq:Psi_SIS})
\begin{align}
\psi(\bs{\theta}_i) = \alpha_0 ||\bs{\theta}_i||\,.
\end{align}
Expanding this expression to linear order in $v_o/c$, one finds
\begin{align}
\psi(\bs{\theta}_i) = \psi'_i + P_i \frac{v_o}{c}\,,
\end{align}
where
\begin{align}
\psi_i' & = \alpha_0' ||\bs{\t{\theta}_i'}||\,,\\
P_i & = \frac{1}{||\bs{\t{\theta}_i'}||} (\alpha_0' \bs{\T}_i \cdot \bs{\t{\theta}'_i} + A_0 || \bs{\t{\theta}'_i}||^2)\,, \label{eq:P}
\end{align}
where $\bs{\T_i}$ and $A_0$ were defined in Eq.\,\eqref{eq:T} and \eqref{eq:A_0}. It is affected directly by the peculiar velocity bias on the images and indirectly by the bias on $\alpha_0$, which comes from the bias on the distances and on the velocity dispersion. As such, it is sensitive to the peculiar velocities of the source, lens and observer.

\subsection{The time delay and Hubble constant}\label{sec:Time_Delay_Hubble_Constant}

At this point, all necessary contributions to the bias on the time delay have been computed and we expand the right-hand side of Eq.\,\eqref{eq:Time_Delay} to first order in $v_o/c$, while the left-hand side is invariant, up to $\mathcal{O}(v_o^2/c^2)$ (see Eq.\,\eqref{eq:Time_Dilation}). We get
\begin{widetext}
\begin{align}
c \Delta t_{ij}\simeq c \Delta t_{ij}' = (1+z_l')  \frac{d[z_l'] d[z_s']}{ d[z_l',z_s']} \l\{ \l[ \frac{(\bs{\theta}_i' - \bs{\beta}'')^2}{2}- \frac{(\bs{\theta}_i' - \bs{\beta}'')^2}{2} \r] - [\psi_i' - \psi_j']\r\} + c \Delta T_{ij} \frac{v_o}{c}\,,\label{eq:Biased_Time_Delay}
\end{align}
where the (distance) time-delay bias is given by
\begin{align} 
c \Delta T_{ij} & = (1+z_l') \frac{d[z_l'] d[z_s']  }{d[z_l',z_s']}  \Big[   (\bs{\t{\theta}}_i'-\bs{\t{\beta}_i''} ) (\bs{\T_i}-\bs{B''_i}) - ( \bs{\t{\theta}}_j' -\bs{\t{\beta}_j''})(\bs{\T_j}-\bs{B_j''})  -(P_i - P_j)  \Big] \nonumber \\
& \quad +  \frac{1+z_l'}{d^2[z_l',z_s']} \l[  Z_L d[z_l'] d[z_l',z_s'] d[z_s'] + d[z_l'] d[z_l',z_s'] D_S - d[z_l'] D_{LS} d[z_s'] + D_L d[z_l', z_s'] d[z_s']\r] \nonumber \\
& \quad \times \l[ \hat{\phi}(\bs{\t{\theta}_i'}, \bs{\t{\beta}_i''}) -\hat{\phi}(\bs{\t{\theta}_j'}, \bs{\t{\beta}_j''})\r] \,,\label{eq:Tij}
\end{align}
\end{widetext}
which can be computed directly from observables, following the steps provided in subsections \ref{sec:Time_Dilation}-\ref{sec:Lensing_Potential}. In particular, it can be computed directly from the observed redshifts $z_l'$, $z_s'$, their associated distances (Eqs. \eqref{eq:dlp}-\eqref{eq:dlsp}), images $\bs{\t{\theta}_i'}$, the reconstructed source angle $\bs{\t{\beta}''}$ via Eq.\,\eqref{eq:betapp}, the angle biases $\bs{\T_i}$, $\bs{B_i''}$ defined in Eq.\,\eqref{eq:T} and Eq.\,\eqref{eq:Bpp}, the lensing potential biases $P_i$ defined in Eq.\,\eqref{eq:P}, and the distance biases $D_L$, $D_S$ and $D_{LS}$ defined in Eq.\,\eqref{eq:DL}, \eqref{eq:DS} and \eqref{eq:DLS}. One can recognize the contributions coming from the bias on angles in the first line, together with the lensing potential. Those are directly affected by the peculiar velocity of the observer, and indirectly affected by the peculiar velocities of the source and lens through the lens model. The second line is due to the direct bias on redshift and angular diameter distances as a function of observed redshift from the peculiar velocity of the source, lens and observer. Solving Eq.\,\eqref{eq:Biased_Time_Delay} for $H_0$, which appears in the angular diameter distance ratio, one gets
\begin{align}
H_0 & =\underbrace{\frac{\chi[z_l'] \chi[z_s']}{ \chi[z_l',z_s']} \frac{\l[ \hat{\phi}(\bs{\theta_i'},\bs{\beta_i''})- \hat{\phi}(\bs{\theta_j'},\bs{\beta_j''}) \r]}{\Delta t_{ij}' }}_{=H_0'} \l( 1 + \frac{c \Delta T_{ij} }{c \Delta t_{ij}' }\frac{v_o}{c}\r) \equiv H_0'\l (1 + \frac{\Delta H_0}{H_0'}\r)\,,
\end{align}
with
\begin{empheq}[box=\fbox]{equation}
\frac{\Delta H_0}{H_0'} = \frac{\Delta T_{ij}}{\Delta t_{ij}'} \frac{v_o}{c}\,.\label{eq:H0_bias}
\end{empheq}
Equation \eqref{eq:H0_bias} together with Eq.\,\eqref{eq:Tij} are the main results of this work. For a given pair of images with measured $\{ \Delta t_{ij}', z_l', z_s', \bs{\t{\theta}'_i},\bs{\t{\theta}'_j} \}$ and given peculiar velocities, one can compute the corresponding bias $\Delta H_0/H_0'$ as a function of $H_0$. This is because $\Delta T_{ij}$ given in Eq.\,\eqref{eq:Tij} is inversely proportional to $H_0$. Alternatively, one can compute $\Delta H_0$ independently of $H_0$ to first order in $v_o$ since the ratio $H_0'/H_0$ which would appear on the right-hand side of Eq.\,\eqref{eq:H0_bias} only brings second order corrections. Throughout the manuscript, we take $H_0=70$ km s$^{-1}$ Mpc$^{-1}$. In the next section, we apply these findings to the seven lenses of TDCOSMO \cite{Wong:2019kwg, Shajib:2020, Millon2020}. It should be noted also that with this definition, a positive $\Delta H_0$ implies that $H_0'$ is an underestimation of $H_0$. Therefore, a relatively high $H_0'$ could be explained by a negative $\Delta H_0/H_0'$.

\section{Results}\label{sec:Results}

\begin{widetext}
\begin{figure}
    \centering
    \includegraphics[width=0.7\textwidth]{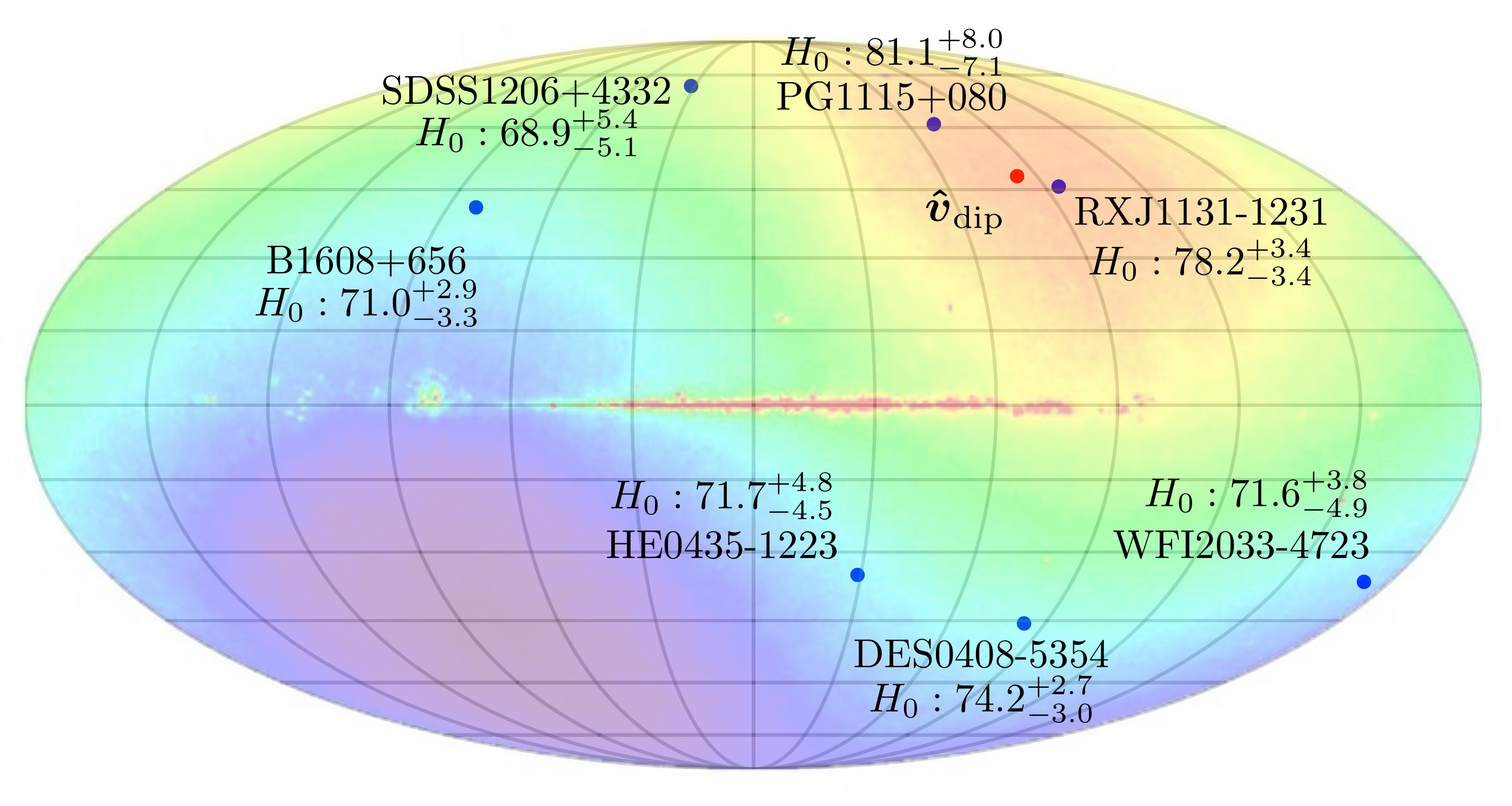}
    \caption{Blue dots indicate the sky position in galactic coordinates of the 7 lenses of H0LiCOW together with their corresponding estimation of $H_0$ (in km s$^{-1}$ Mpc$^{-1}$), extracted from \cite{Wong:2019kwg,Shajib:2020}. Their sky positions are given in Table \ref{tab:parameters}. We superimpose the CMB temperature map from WMAP \cite{WMAP}, where the monopole has been removed, leaving the dipole apparent, together with contamination from the galactic plane. The red dot indicates the direction of the velocity obtained from the CMB dipole $\bs{v}\e{dip}$. The two lenses RXJ1131-1231 and PG1115+080 have the two lines of sight which are best aligned with the CMB dipole, with $\cos(\theta_{cm}')>0.96$. Coincidentally, they also have the lower lens redshift and give the highest values of $H_0$: $78.2^{+3.4}_{-3.4}$ km s$^{-1}$ Mpc$^{-1}$ and $81.1^{+8.0}_{-7.1}$ km s$^{-1}$ Mpc$^{-1}$, respectively. This was pointed out in \cite{Krishnan:2021jmh}. As we shall see, the peculiar velocity of the observer should be about 8 times larger than $v\e{dip}$, i.e. $\sim3000$km s$^{-1}$ pointing in the opposite direction of $\bs{v}\e{dip}$ to explain alone these relatively high $H_0$ values (See Fig.\, \ref{fig:vo_variation}). The CMB dipole in \textit{celestial} coordinates is $\bs{\hat{v}}\e{dip} \simeq ( -7^\circ,167^\circ)$, which is well aligned with the Earth's equator. In this sense, North or South hemisphere sky surveys are nearly as orthogonal as they can be from the CMB dipole. }
    \label{fig:Sky_Lenses}
\end{figure}
\end{widetext}

In this section, we quantify what is the relative bias on $H_0$ from the peculiar velocities of the observer, lens and source for the seven lenses of H0LiCOW. We first consider our results with expected peculiar velocities, before considering what happens for larger peculiar velocities. We get estimations of the Hubble constant $H_0'$ which vary between $47$ km s$^{-1}$ Mpc$^{-1}$ and $112$ km s$^{-1}$ Mpc$^{-1}$. Since the model is relatively crude, we do not expect to make a competitive inference of the Hubble constant. The SIS model is spherically symmetric and fixes the logarithmic slope of the mass profile to $\gamma_l = 2$. This of course does not contain enough azimuthal and radial degrees of freedom to represent accurately massive elliptical galaxies. However, we expect this model to be sufficient to capture the leading contributions to a bias on $H_0$ from peculiar velocities. In Fig.\,\ref{fig:Sky_Lenses}, we plot the sky distribution of the 7 lenses. Two are well aligned with the velocity $\bs{\hat{v}}\e{dip}$, namely RXJ1131$-$1231 and PG1115$+$080. These two systems coincidentally also happen to have the lowest lens redshifts and the highest inference of the Hubble constant. 
\begin{table*}[htbp] 
\centering
\renewcommand{\arraystretch}{1.5}
\begin{tabular}{l|cccccccc} 
$N$ & Lens system & $z_l'$ & $z_s'$ & $\bs{\hat{n}'}$ $(l',b')$ [$^\circ$] & $\cos (\theta_{cm}')$ & $N\e{images}$ & $\Delta H_0/H_0'$ [$\%$]& Reference \\ \hline 
1 & B1608+656 & 0.6304 & 1.394 & (98.339, 40.891) & 0.000706 & 4 & 0.0006 & \cite{Suyu2010, Jee2019} \\ 
2 & RXJ1131-1231 & 0.295 & 0.654 & (-85.573,45.888) & 0.991526 & 4 & 1.1353 & \cite{Suyu2014,Chen2019} \\ 
3 & HE0435-1223 & 0.4546 & 1.693 & (-150.934, -35.060) & -0.115625 & 4 & -0.2153 & \cite{Wong2017, Chen2019} \\ 
4 & SDSS1206+4332 & 0.745 & 1.789 & (148.991, 71.244) & 0.615891 & 2 & 0.2324 & \cite{Birrer2019} \\
5 & WFI2033-4723 & 0.6575 & 1.662 & (-7.585, -36.556) & -0.429394 & 4 & -0.5008 & \cite{Rusu2019} \\
6 & PG1115+080 & 0.311 & 1.722 & (-110.113, 60.644) & 0.966824 & 4 & 1.1461 & \cite{Chen2019} \\
7 & DES0408-5354 & 0.597 & 2.375 & (-96.447, -45.304)& -0.0620664 & 3 & -0.0590 &\cite{Shajib:2020} \\
\end{tabular} 
\caption{This table contains the system number, their lens systems with observed lens and source redshift, optical axis directions in galactic coordinates, projection of the line of sight along the peculiar velocity of the observer $\bs{v}\e{dip}$ and the number $N\e{images}$ of effective images that can be used for time-delay cosmography per system. There is also a column indicating the relative bias $\Delta H_0/H_0'$ generated by the observer's peculiar velocity $\bs{v}\e{dip}$ alone. The latter is averaged over the non redundant pairs of images.
\label{tab:parameters}}
\end{table*}

\begin{figure}
    \centering
    \includegraphics[width=0.85\textwidth]{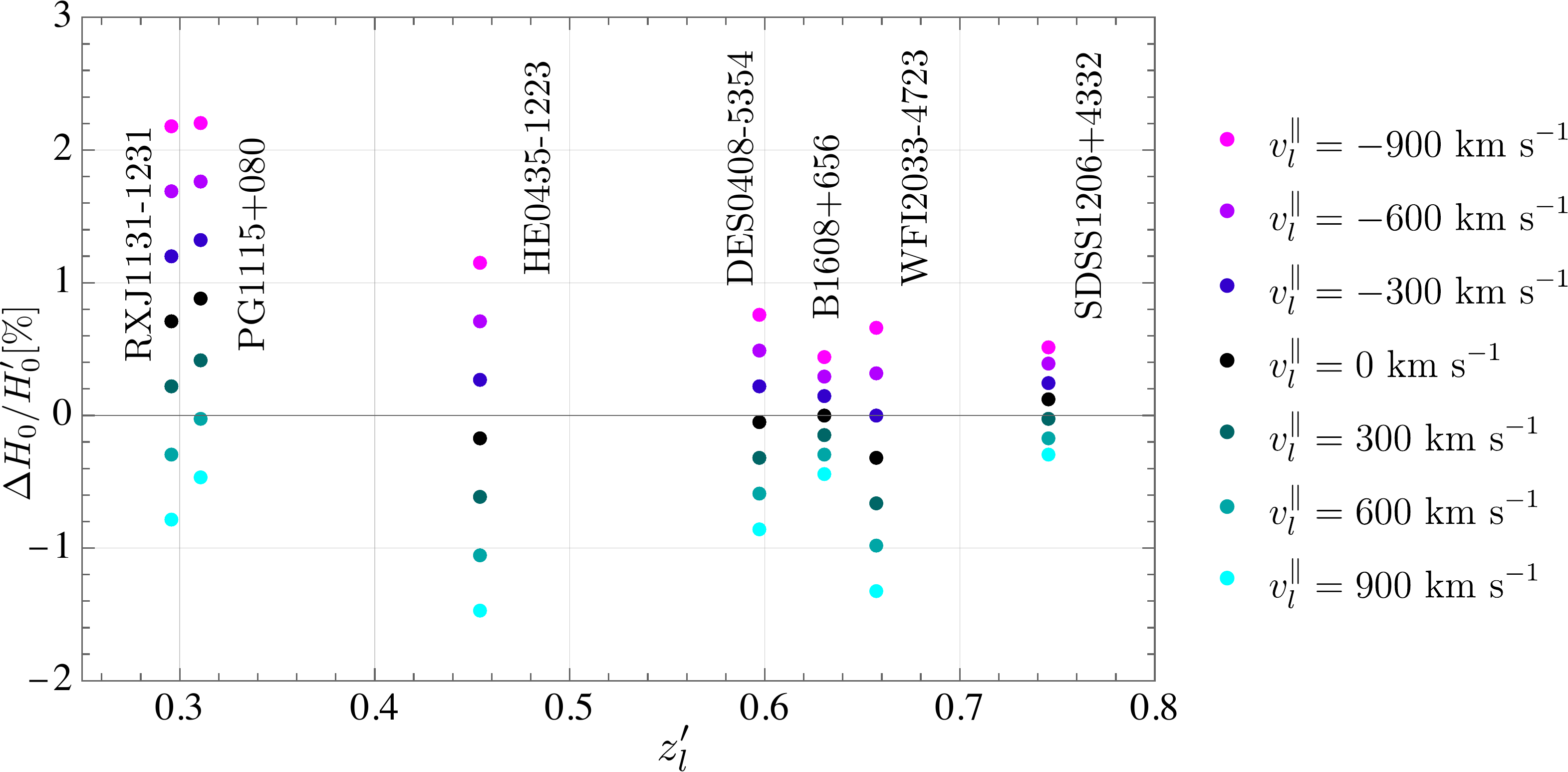}
    \caption{In this plot, we show the normalized bias on $H_0$, for $v_o$ extracted from the entirely kinematic interpretation of the CMB and vary $v_l^\parallel$, $v_s^\parallel \in \{ 0,\pm 300,\pm 600, \pm 900\}$ km s$^{-1}$. For these plots, we assumed that $\sigma_v$ can be measured independently from the peculiar velocities. This implies that we set the lens parameter bias $S_v =0$, instead of using the expression for $S_v$ given in Eq.\,\eqref{eq:S_v}. While varying $v_s^\parallel$ does change the bias on $H_0$, the change is much smaller than that of $v_l^\parallel$ and the points with different $v_s^\parallel$ appear to coincide on this plot. In this case, the amplitude of the bias is bounded by $2.5\%$.}
    \label{fig:vl_variation_Sv_is_0}
\end{figure}

First, we compute the bias generated by the peculiar velocity of the observer, assuming that it is known from the entirely kinematic interpretation of the CMB dipole. That corresponds to $v_o=369.82$ km s$^{-1}$ toward $(264.021^\circ, 48.253^\circ )$ in galactic coordinates. Then, we vary the source and lens peculiar velocities projected on the line of sight in the set $\{0,\pm 300, \pm 600,\pm 900\}$ km s$^{-1}$, which spans the expected peculiar velocity amplitudes from simulations \cite{Bertschinger1991} and from observations \cite{Girardi:1998rm}. We do this for two different cases. In the first case, we assume that $\sigma_v$ can be measured independently of the peculiar velocities, from the lens galaxy emission lines' width. In this case, only the peculiar velocity of the lens on top of the peculiar velocity of the observer changes the bias on $H_0$ in a way that can be seen in Fig.\,\ref{fig:vl_variation_Sv_is_0}, where we plot $\Delta H_0/H_0'$ as a function of the lens redshift. In this plot and in the following, $\Delta H_0/H_0'$ is actually the average over the nonredundant image pairs available. In practice, for a given system, some time delays are more precisely measured than others and therefore a weighted average may be more sensible to compute the relative bias on $H_0$. The source peculiar velocity only affects the bias subdominantly. The bias for one lens is bounded by $2.5\%$ and the bias generated by the observer's peculiar velocity alone is bounded by $1\%$. 

\begin{figure}
    \centering
    \includegraphics[width=0.85\textwidth]{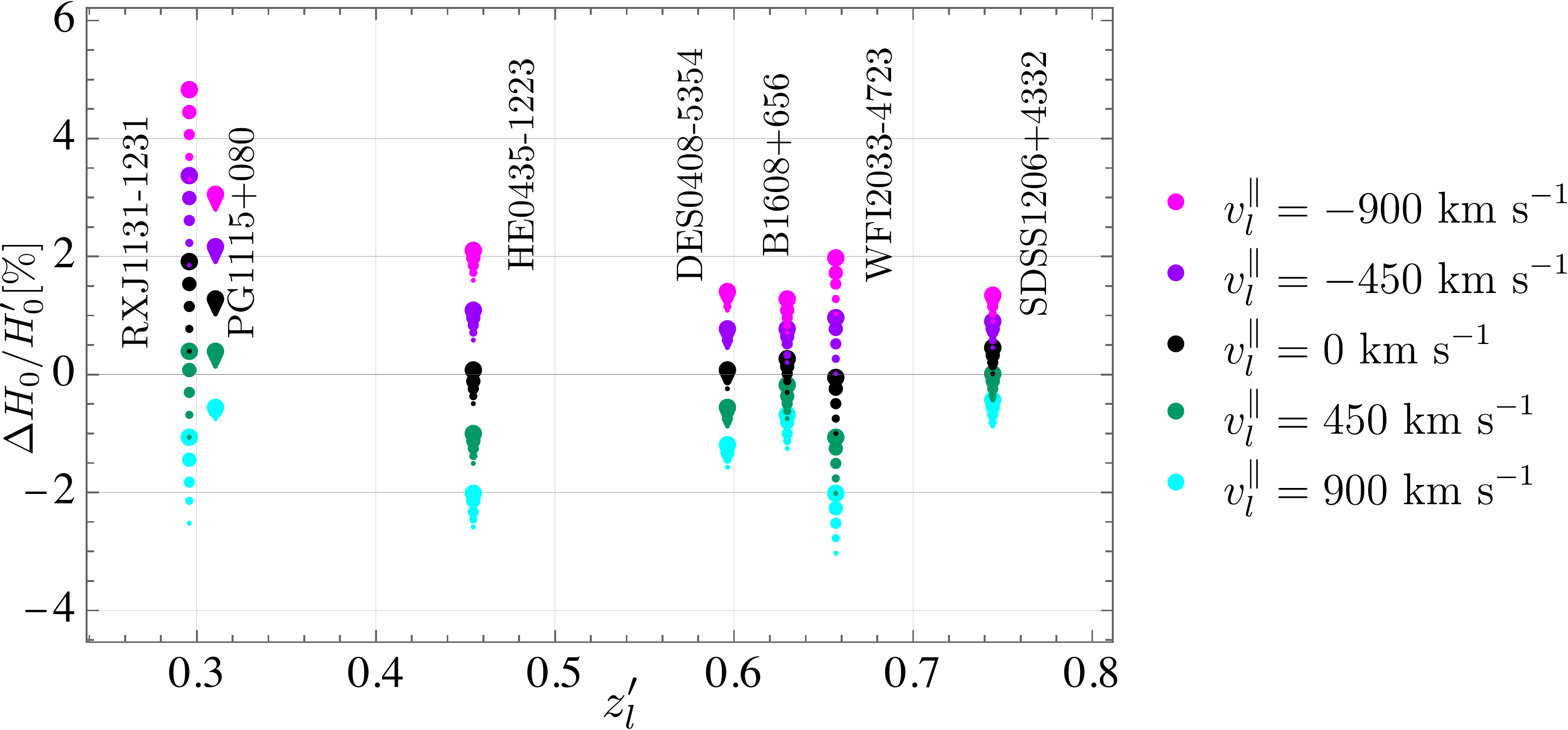}
    \caption{In this plot, we show the normalized bias on $H_0$, for $v_o$ extracted from the entirely kinematic interpretation of the CMB and vary $v_l^\parallel$, $v_s^\parallel \in \{ 0,\pm 450, \pm 900\}$ km s$^{-1}$. Larger dots indicate larger source peculiar velocities $v_s^\parallel$. For this plot, we assumed that $\sigma_v$ is extracted from the observed Einstein angle, as outlined in Sec.\,\ref{Sec:Peculiar_velocity_bias}. This implies that $S_v$ is calculated using Eq.\,\eqref{eq:S_v}, contrary to Fig.\,\ref{fig:vl_variation_Sv_is_0}, where it was set to zero. While varying $v_s^\parallel$ does change the bias on $H_0$, the change is smaller than that of $v_l^\parallel$. Note that this is different to the situation presented in Fig.\,\ref{fig:vl_variation_Sv_is_0}, where the velocity of the source affects less the bias on $H_0$. The largest bias appears for lens and source peculiar velocities which are antialigned. In this case, the amplitude of the bias on the Hubble constant can reach $5\%$. The peculiar velocity of the observer alone gives an amplitude bias which is bounded by $1.2\%$. }
    \label{fig:vs_variation_with_Sv}
\end{figure}

In the second case, we assume that $\sigma_v'$ is extracted from the measurement of the Einstein angle, as outlined in Sec.\,\ref{Sec:Peculiar_velocity_bias}. In that case, both the lens and the source peculiar velocities give significant changes to the bias on $H_0$, as can be seen in Fig.\,\ref{fig:vs_variation_with_Sv}, where we plot $\Delta H_0/H_0'$ as a function of the lens redshift $z_l'$. 
In this case, the bias $\Delta H_0/H_0'$ for a single lens is bounded by $5\%$ for these seven lenses. The effect of the peculiar velocity of the observer alone, as extracted from the CMB dipole, is bounded by $1.2\%$. This shows how the effect of $v_o/c= \mathcal{O}(10^{-3})$ can give an order of magnitude larger bias, as the bias piles up from different observables. In table \ref{tab:biases}, we give the maximal relative bias on each quantity that enters Eq.\,\eqref{eq:Tij} from the velocity of the observer set to $v\e{dip}$ for each of the seven systems of TDCOSMO. Combining the seven lenses, we find that the bias generated on $H_0$ by the observer's peculiar velocity is of order $0.25\%$. Assuming that the lens and source peculiar velocities are normally distributed around zero with standard deviation $300$\,km\,s$^{-1}$, one finds that this results in an additional random uncertainty which can reach $1.00\%$ for a single lens. It combines to a $0.24\%$ random uncertainty for the seven lenses of TDCOSMO. This uncertainty is expected to drop to zero for a higher number of systems.

\begin{table*}[htbp] 
\centering
\renewcommand{\arraystretch}{1.7}

\begin{tabular}{l|c|ccc|c|c|c|c|c}
\multirow{2}{*}{N} & \multirow{2}{*}{Lens system} & \multicolumn{3}{c|}{Angular diameter distances} & \begin{tabular}[c]{@{}c@{}}Deflection \\ angle\end{tabular} & \begin{tabular}[c]{@{}c@{}}Lensing \\ potential\end{tabular} & \begin{tabular}[c]{@{}c@{}}SIS velocity \\ dispersion\end{tabular} & \begin{tabular}[c]{@{}c@{}}Source \\ position angle\end{tabular} & \begin{tabular}[c]{@{}c@{}}Images \\ position angle\end{tabular} \\ \cline{3-10} 
 &  & \multicolumn{1}{c|}{$\frac{D_L}{d[z_l']} \frac{v\e{dip}}{c}$ [$\%$]} & \multicolumn{1}{c|}{$\frac{D_S}{d[z_s']} \frac{v\e{dip}}{c}$ [$\%$]} & $\frac{D_{LS}}{d[z_l',z_s']} \frac{v\e{dip}}{c}$ [$\%$] & $\frac{A_0}{\alpha_0} \frac{v\e{dip}}{c}$ [$\%$] & $|\frac{P_i}{\psi_i'}| \frac{v\e{dip}}{c}$ [$\%$] & $\frac{S_v}{\sigma_v'} \frac{v\e{dip}}{c}$ [$\%$] & $ \frac{||\bs{B''_i}||}{||\bs{\t{\beta}''_i}||}\frac{v\e{dip}}{c}$ [$\%$] & $ \frac{||\bs{\Theta_i}||}{||\bs{\t{\theta}'_i}||}\frac{v\e{dip}}{c}$ [$\%$] \\ \hline
1 & B1608+656 & \multicolumn{1}{c|}{0.0001} & \multicolumn{1}{c|}{0.00001} & -0.0001 & 0.0001 & 0.0003 & 0.0001 & 1.1011 & 0.2846 \\
2 & RXJ1131-1231 & \multicolumn{1}{c|}{0.3739} & \multicolumn{1}{c|}{0.1325} & -0.1051 & 0.1221 & 0.2107 & 0.1799 & 0.6470 & 0.0986 \\
3 & HE0435-1223 & \multicolumn{1}{c|}{-0.0259} & \multicolumn{1}{c|}{0.0006} & 0.0161 & -0.0143 & 0.0282 & -0.0149 & 2.3707 & 0.1293 \\
4 & SDSS1206+4332 & \multicolumn{1}{c|}{0.0663} & \multicolumn{1}{c|}{0.-0.0064} & -0.0925 & 0.0759 & 0.1516 & 0.0810 & 0.3607 & 0.0880 \\
5 & WFI2033-4723 & \multicolumn{1}{c|}{-0.0569} & \multicolumn{1}{c|}{0.0015} & 0.0627& -0.0530 & 0.1054 & -0.0571 & 1.2443 & 0.1443 \\
6 & PG1115+080 & \multicolumn{1}{c|}{0.3431} & \multicolumn{1}{c|}{-0.0066} & -0.1294 & 0.1191 & 0.2353 & 0.1210 & 5.5110 & 0.1182 \\
7 & DES0408-5354 & \multicolumn{1}{c|}{-0.0095} & \multicolumn{1}{c|}{0.0021} & 0.0094 & -0.0076 & 0.0154 & -0.0075 & 0.9690 & 0.2024
\end{tabular}
\caption{We give the relative biases on each quantity assuming that $v_l=0=v_s$ and that the observer has a peculiar velocity of amplitude $||\bs{v}\e{dip}||=369.82$ km s$^{-1}$, as extracted from the entirely kinematic interpretation of the CMB dipole. Most of the biases are below the percent level. When there are several values for a single system, for example for $P_i/\psi_i'$, $||\bs{B''_i}||/||\bs{\t{\beta}''_i}||$, $||\bs{T_i}||/||\bs{\t{\theta}'_i}||$, we give only the result for the image which maximizes the bias.
}\label{tab:biases}
\end{table*}

\begin{figure}
    \centering
    \includegraphics[width=0.85\textwidth]{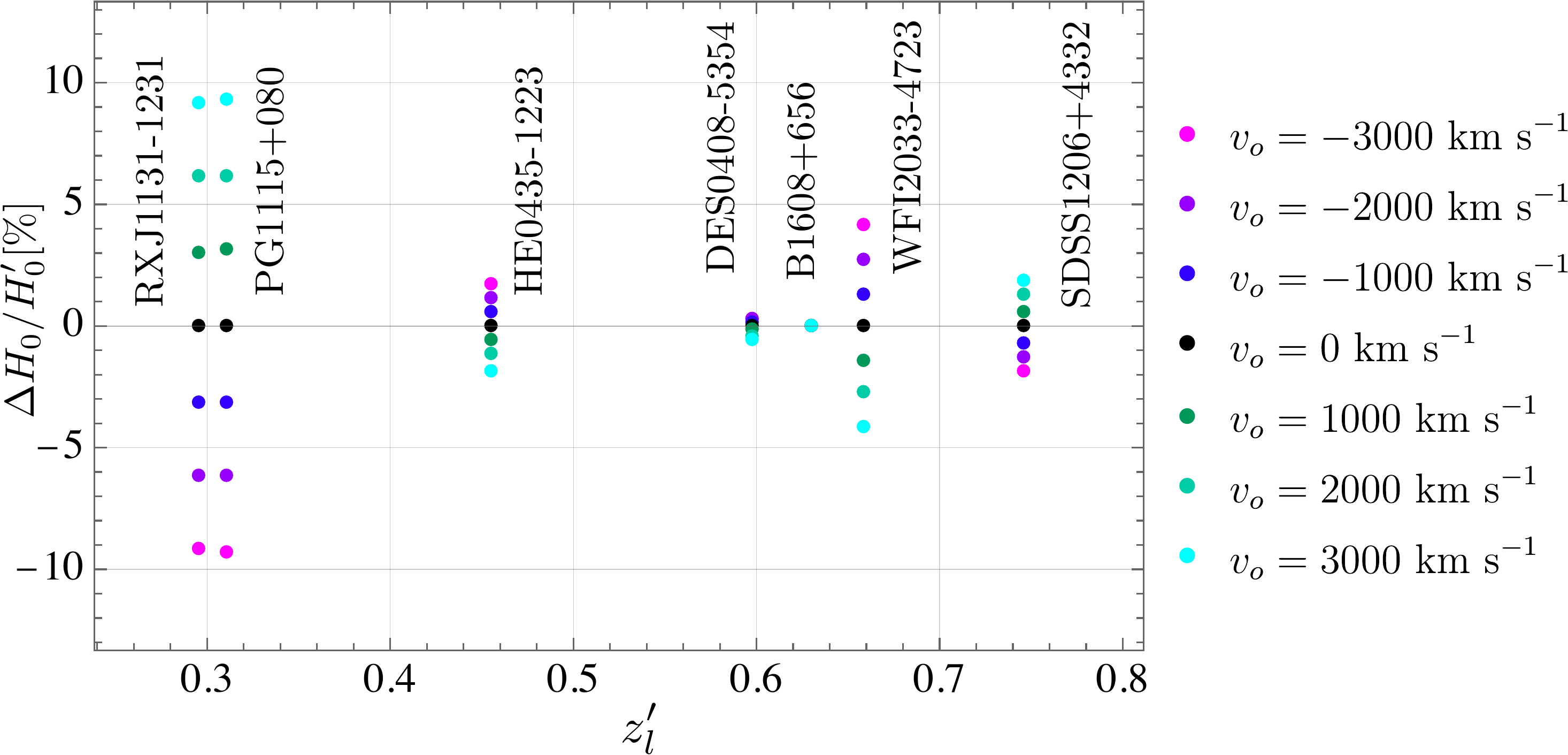}
    \caption{In this plot, we show the relative bias $\Delta H_0/H_0'$ on $H_0$ for an observer with peculiar velocity $v_o\in \{0,\pm1000,\pm2000,\pm3000\}$ km s$^{-1}$ as a function of the observed lens redshift $z_l'$. Peculiar velocities of the order of $\pm3000$ km s$^{-1}$ are required to bias the Hubble constant to the order of 10$\%$ for the systems which are best aligned with $\bs{v}\e{dip}$, which are RXJ1131-1231 and PG1115+080. This corresponds to more than 8 times more than $\bs{v}\e{dip}$. One might be intrigued by the sign of the bias for these two lenses. It turns out that a negative velocity $-3000$ km s$^{-1}$ is required, to bias $H_0'$ to $\sim 10 \%$ higher than $H_0$. In this sense, the higher $v_o$ expected from number count dipoles works against lowering the Hubble constant extracted from the two low redshift lenses RXJ1131-1231 and PG1115+080. Here, peculiar velocities of the lens and source are set to zero and $\sigma_v$ is assumed to be extracted from the observed Einstein angle, as outlined in Sec.\,\ref{Sec:Peculiar_velocity_bias}. The system B1608+656 at lens redshift $z_l' \simeq 0.63$ is nearly not affected by the boost because it is quasi-orthogonal to the CMB dipole direction. The situation would change if the direction of $\bs{v}_o$ was also varied. }
    \label{fig:vo_variation}
\end{figure}

\begin{figure}
    \centering
    \includegraphics[width=0.85\textwidth]{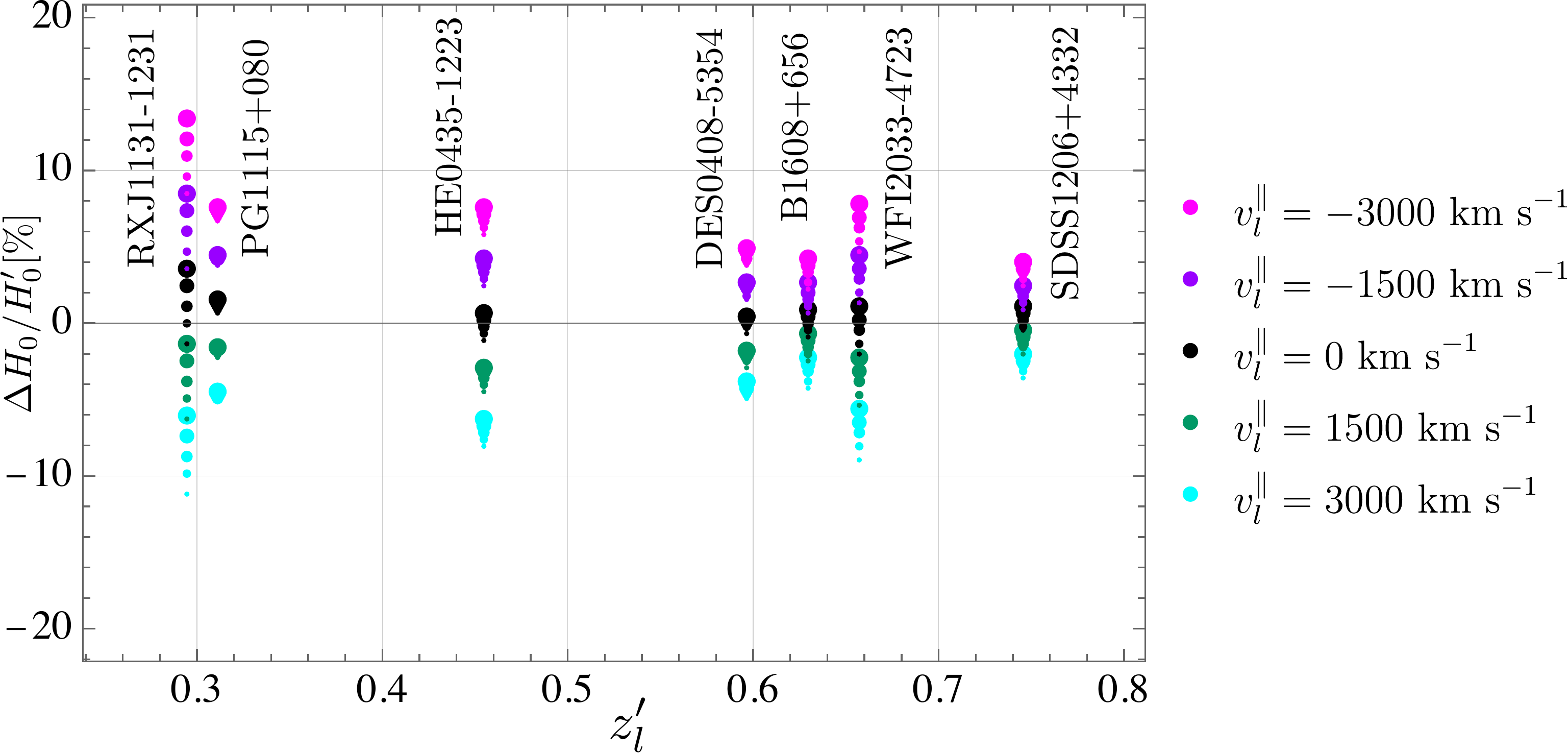}
    \caption{We plot the bias on $H_0$ from the peculiar velocity of the observer, lens and source. Here $v_o$ is fixed by the entirely kinematic interpretation of the CMB dipole, i.e.\,$v_o = 369.82$ km s$^{-1}$. The lens and source peculiar velocities are allowed to vary in $\{ 0, \pm 1500,\pm 3000\}$ km s$^{-1}$. Larger dots indicate larger source peculiar velocities $v_s^\parallel$. Here $\sigma_v$ was computed from the observed Einstein angle, as outlined in Sec.\,\ref{Sec:Peculiar_velocity_bias}. In this case, $S_v\neq 0$. This corresponds to what has mostly been done in practice in \cite{Wong:2019kwg,Shajib:2020}. In this case, the velocity of the lens influences significantly the bias on $H_0$ and the source peculiar velocity, less so. The largest bias in magnitude appears for the source and lens peculiar velocities which are anti-aligned.}
    \label{fig:vs_large_variation_with_Sv}
\end{figure}

Since the calculation is valid for nonrelativistic velocities, one may push to larger peculiar velocities, as long as $v/c\ll 1$. One may be curious to see what peculiar velocities would be necessary to affect the Hubble constant by $10\%$, which would constitute an important correction in the context of the Hubble tension. We plot the bias from the velocity of the observer for peculiar velocities which vary in $v_o\in \{0,\pm1000,\pm2000,\pm 3000\}$ km s$^{-1}$ in Fig.\,\ref{fig:vo_variation}. Negative peculiar velocities correspond to changing the direction of the peculiar velocity by a rotation of $\pi$. For $v_o=\pm 3000$, the bias for the best-aligned lenses (system RXJ1131$-$1231 and PG1115+080), at observed lens redshift $z_l'\sim 0.3$, reaches $\pm 10\%$. It is intriguing that the number count dipole measurements point to higher $\bs{v}_o$, which argues in favor of positive $\Delta H_0$. This suggests that if the peculiar velocity of the observer is higher than expected, even by a factor of $10$, then the estimation of the Hubble constant by the H0LICOW collaboration is rather an underestimation of $H_0$, which would enhance the tension. For $v_o =0$, which corresponds to a comoving observer, the bias vanishes, as expected. The bias changes in different directions and with different amplitudes for different systems. This depends on the sign and value of $\cos(\theta_{cm}')$ together with the lens and source redshifts, which are given in Table\,\ref{tab:parameters}.
Finally, we play the same game with peculiar velocities of the lens and source. We vary them in $\{0,\pm1500,\pm3000\}$ km s$^{-1}$. The assumption on $\sigma_v$ determines how less important $v_s^\parallel$ matters compared to $v_l^\parallel$ for the bias on $H_0$. Since in practice, $\sigma_v$ is extracted from the Einstein angle, we plot what happens in that case in Fig.\,\ref{fig:vs_large_variation_with_Sv}. These large peculiar velocities, which are expected to be rare, can bias $H_0$ by more than $10\%$. However, the directions of these lense and source peculiar velocity would have to conspire to always bias $H_0$ in the same direction, which is unexpected in isotropic cosmologies. 

\section{Conclusion}\label{sec:Conclusion}

In this work, we quantified the effects of peculiar velocities of the lens, source and observer on the determination of the Hubble constant from time-delay cosmography, carefully taking into account all boost effects on the observables and their repercussion on the lens model. We showed in detail how to compute the bias, given peculiar velocities and assuming that the lens is well described by a singular isothermal sphere. Even if this model alone does not allow for more than two images per lensed quasar, and gives crude estimates of $H_0$, we expect this model to be sufficient to capture the leading effects of peculiar velocities on current time-delay cosmography experiments. For the observer's peculiar velocities fixed to $||\bs{v}\e{dip}|| = 369.82$ km s$^{-1}$, as extracted from the entirely kinematic interpretation of the CMB dipole, the bias on $H_0$ is, at most, of the order of the percent level for a single lens. The sign and amplitude of the bias depends on the direction of the observed lens center of mass, which is captured by $\cos (\theta_{cm}')$. The bias on $H_0$ from the observer's peculiar velocity for the combined seven lenses, which span different corners of the sky is of $0.25\%$. These cancellations for the observer's peculiar velocity require an isotropic distribution of lensed quasars, which may be jeopardized by the specific footprint of time-domain surveys. This is however mitigated by the fact that the CMB dipole points to celestial declination $-7^\circ$, which is close to the Earth's equator. In this sense, North or South hemisphere surveys are nearly as orthogonal as they can be from the observer's peculiar velocity. 

If one includes the effect of the lens and source peculiar velocities projected on the line of sight, up to $|v_l^\parallel|$, $|v_s^\parallel| \leq 900$ km s$^{-1}$, then the effect reaches at most $5 \%$. The sign of these contributions depends entirely on the sign and amplitude of these peculiar velocities, which vary from one system to another and may be expected to cancel out between a source and another, for a sufficiently high number of systems. Assuming that the lens and source peculiar velocities are normally distributed around zero with standard variation of $300$ km s$^{-1}$, we found that these generate a random uncertainty on $H_0$, which can reach $1.00\%$ for a single lens and which combines to $0.24\%$ for the seven systems. We also found that the way that the lens model parameter, i.e.\,the velocity dispersion $\sigma_v$ is determined, affects how subdominant the source peculiar velocities are in the Hubble constant bias. If one can determine the velocity dispersion independently of the peculiar velocities of the observer and lens, from spectroscopic measurements, then the bias from peculiar velocities on the Hubble constant is reduced. This can bring the bias from $5\%$ to $2.5\%$ in the most extreme cases with $v_l^\parallel=-900$ km s$^{-1}$ antialigned with $v_s^\parallel = 900$ km s$^{-1}$. A measurement of the peculiar velocities by an alternative distance measurement would allow to correct for the bias and remove this source of random uncertainties. In this regard, the redshift difference between two images of a strongly lensed source was suggested as a probe of the source peculiar velocity \cite{Molnar:2013}.

Finally, we studied what peculiar velocities are required to bias the Hubble constant determination to the order of $10\%$. We found that peculiar velocities projected on the line of sight of the order of $3000$ km s$^{-1}$ would do the job. This can be cumulated between the observer, the source and the lens. This requires unexpectedly large peculiar velocities. Coincidentally, the two systems which are best aligned with $ \bs{v}\e{dip}$ are also the ones which give the higher $H_0$ estimates in H0LiCOW \cite{Wong:2019kwg}. It is interesting, in light of the number count experiments which favor a larger $||\bs{v}_o||$, that a larger observer peculiar velocity works against resolving the Hubble tension since it would imply that the H0LiCOW collaboration rather underestimates $H_0$ for these two lenses, which already give the highest $H_0$ estimates. In other words, lowering the Hubble constant estimates for these two lenses requires an observer velocity which goes in the opposite direction of the CMB, with an amplitude roughly $8$ times larger than $||\bs{v}\e{dip}||$. Future biased estimations of the Hubble constant could also be expected if one observes systems consistently in the same hemisphere aligned with the observer's peculiar velocity. Even more so if the observer's peculiar velocity is larger in magnitude than one expects from the entirely kinematic interpretation of the CMB dipole.

The small number of sources ($\mathcal{O}(10)$) implies that cancellations over many different sources which are distributed isotropically may be spoiled by shot noise. If it is clear that these large peculiar velocities are rare in $\Lambda$CDM, to rule them out would require distance estimates, which combined with redshifts, can be used to constrain the lens and source peculiar velocities. To affect the Hubble constant consistently over many sources would require large bulk flows of sources which are not expected in homogeneous and isotropic cosmologies. In $\Lambda$CDM, one expects the bulk flow velocity of sources on a sphere centered on the observer to decay with increasing radius. It should be noted that several anomalies have been pointed out in such convergence to the Hubble flow \cite{Aluri:2022hzs,Kashlinsky:2010, Kashlinsky:2008,Gunn:1988,Migkas:2021zdo} on scales which can reach up to $800$ Mpc. For example, these large peculiar velocities could be expected for an observer who is offset from the center of an ultra-large void, which was studied in \cite{Cai:2022dov} and proposed as a solution to the cosmic dipole tension. In this scenario, these large peculiar velocities could be interpreted as artifacts from working with the wrong background equations of motion. 

Finally, we conclude that peculiar velocities of the observer, source and lens play a significant role in time-delay cosmography, if one is after percent precision on the Hubble constant. It seems difficult to accommodate a larger observer's peculiar velocity, as suspected from radio source and quasar number counts, as a simultaneous explanation for the bias toward higher $H_0$ from time-delay cosmography. Future independent constraints on the peculiar velocities of the lenses, sources and observer could help to constrain the Hubble constant to percent precision using time-delay cosmography.

\section{Acknowledgments}

We would like to thank Pierre Fleury, Eric Linder and Sebastian Von Hausegger for interesting discussions and Aymeric Galan and Simon Birrer for valuable feedback on a preliminary version of this work. C.D.\,and T.B.\,are supported by ERC Starting Grant \textit{SHADE} (grant no.\,StG 949572). M.M.\,acknowledges the support of the Swiss National Science Foundation (SNSF) under grant P500PT\_203114. T.B.\,is further supported by a Royal Society University Research Fellowship (grant no.\,URF$\backslash$ R1$\backslash$180009).

\bibliographystyle{unsrt}
\bibliography{references}

%%%%%%%%%%%%%%%%%%%%%%%%%%%%%%%%%%%%%%%%%%%%%%%%%%

%%%%%%%%%%%%%%%%% APPENDICES %%%%%%%%%%%%%%%%%%%%%

\appendix

\section{Rotation angle}
\label{app:Rotation_angle}
The rotation angle $\delta'$ serves to translate the coordinates in the observation frame for a moving observer to the calculation frame. The rotation angle $\delta$ serves to transform these back to the observation frame of a comoving observer. The rotation angle $\delta'$ depicted in Fig.\,\ref{fig:Coordinate_System} can be obtained from the lens' center of mass vectors $\bs{\hat{n}'}$ and the vector $\bs{\hat{N}'} = (122.932^\circ, 27.128^\circ)$ in galactic coordinates, which points in the direction of the Earth's North pole in J2000. The vector $\bs{\tilde{\theta}_y'} =(\t{y}_1', \t{y}_2', \t{y}_3')$
is the projection of the North pole direction $\bs{\hat{N}'}$ in the plane orthogonal to $\bs{\hat{n}'}$, while $\bs{\t{\theta}_x'}$ points East. That is
\begin{align}
\bs{\t{\theta}_y'}= \bs{\hat{N}'} - (\bs{\hat{n}'}\cdot \bs{\hat{N}'}) \bs{\hat{n}'}\,.
\end{align}
The vector $\bs{\hat{\theta}_x'}$ is defined as a vector which is orthogonal both to $\bs{\hat{v}_o}$ and to $\bs{\hat{n}'}$. There are two such vectors which can be obtained by solving the following system for $\bs{\hat{\theta}_x'} = (\hat{x}'_1, \hat{x}'_2, \hat{x}'_3)$
\begin{align}
\bs{\hat{n}'} \cdot \bs{\hat{\theta}_x'} & =0\,,\\
\bs{\hat{v}}_o \cdot \bs{\hat{\theta}_x'} & = 0\,.
\end{align}
The vector $\bs{\hat{\theta}_y'}=(\hat{y}'_1,\hat{y}'_2,\hat{y}'_3 )$ is orthogonal to $\bs{\hat{\theta}_x'}$ and $\bs{\hat{n}'}$ and points toward the positive $\bs{\hat{z}}$ axis, meaning that it is a solution of the following system
\begin{align}
\bs{\hat{\theta}_x'} \cdot \bs{\hat{\theta}_y'} & =0\,,\\
\bs{\hat{n}'} \cdot \bs{\hat{\theta}_y'} & = 0\,,\\
\bs{\hat{\theta}_y'} \cdot \bs{\hat{z}} & > 0\,.
\end{align}
One can compute $\delta'$ in the following way
\begin{align}
\cos \delta' = \bs{\t{\theta}_y'}\cdot \bs{\hat{\theta}_y'}\,.
\end{align}
Since the \textit{comoving} North pole $\bs{\hat{N}}= (\theta_N, \varphi_N)$ and the direction $\bs{\hat{n}}= (\theta_{cm}, \varphi_{cm})$ can be reconstructed using Eqs.\,\eqref{eq:Angle_Transformation_2}-\eqref{eq:Angle_Transformation_phi}, one can repeat these steps to find $\delta$. This defines implicitly the bias $D$ on the rotation angle 
\begin{align}
\delta = \delta' + D \frac{v_o}{c}\,.
\end{align}
Recall that $\delta'$ is the angle between the observation coordinate system spanned by $\{\bs{\t{\theta}'_x},\bs{\t{\theta}'_y} \}$ and a convenient coordinate system $\{\bs{\hat{\theta}'_x},\bs{\hat{\theta}'_y}\}$ as depicted in Fig.\,\ref{fig:Coordinate_System}. This rotation angle is used to determine how the images on the sky appear biased to an observer who has a peculiar velocity $\bs{v}_o$.

\end{document}